\documentclass[prd,aps,preprint, nofootinbib,superscriptaddress]{revtex4-1}
\usepackage{amsmath}
\usepackage{epsfig}

\begin{document}

%\hfill{NPAC-08-24}

\title{The Morphology of the Galactic Dark Matter Synchrotron Emission with Self-Consistent Cosmic Ray Diffusion Models}

\author{Tim Linden}
\email{tlinden@ucsc.edu} \affiliation{Department of Physics, University of California, 1156 High St., Santa Cruz, CA 95064}
\author{Stefano Profumo}
\email{profumo@scipp.ucsc.edu} \affiliation{Department of Physics, University of California, 1156 High St., Santa Cruz, CA 95064}\affiliation{Santa Cruz Institute for Particle Physics, Santa Cruz, CA 95064} 
\author{Brandon Anderson}
\email{anderson@physics.ucsc.edu} \affiliation{Department of Physics, University of California, 1156 High St., Santa Cruz, CA 95064}

\date{\today}

\begin{abstract}
\noindent A generic prediction in the paradigm of weakly interacting dark matter is the production of relativistic particles from dark matter pair-annihilation in regions of high dark matter density. Ultra-relativistic electrons and positrons produced in the center of the Galaxy by dark matter annihilation should produce a diffuse synchrotron emission. While the spectral shape of the synchrotron dark matter haze depends on the particle model (and secondarily on the galactic magnetic fields), the morphology of the haze depends primarily on (1) the dark matter density distribution, (2) the galactic magnetic field morphology, and (3) the diffusion model for high-energy cosmic-ray leptons. Interestingly, an unidentified excess of microwave radiation with characteristics similar to those predicted by dark matter models has been claimed to exist near the galactic center region in the data reported by the WMAP satellite, and dubbed the ``WMAP haze''. In this study, we carry out a self-consistent treatment of the variables enumerated above, enforcing constraints from the available data on cosmic rays, radio surveys and diffuse gamma rays. We outline and make predictions for the general morphology and spectral features of a ``dark matter haze'' and we compare them to the WMAP haze data. We also characterize and study the spectrum and spatial distribution of the inverse Compton emission resulting from the same population of energetic electrons and positrons. We point out that the spectrum and morphology of the radio emission at different frequencies is a powerful diagnostics to test whether a galactic synchrotron haze indeed originates from dark matter annihilation.
\end{abstract}

\maketitle

\section{Introduction}
A compelling paradigm for the particle nature of the dark matter is that of Weakly Interacting Massive Particles, or WIMPs \cite{Bertone:2004pz,Bergstrom:2009ib}. Although the Standard Model of particle physics does not encompass a viable particle dark matter candidate, WIMPs are predicted to exist in several well motivated extensions. These include weak-scale supersymmetry \cite{Jungman:1995df}, models with universal extra dimensions \cite{Hooper:2007qk}, and many others (for reviews see \cite{Bertone:2004pz,Bergstrom:2009ib}). WIMPs are massive particles with masses near the electro-weak scale, and are typically charged under weak interactions. General arguments indicate that WIMPs in thermal equilibrium in the very early universe would {\em freeze-out}, decoupling from the thermal bath when the temperature dropped below a fraction (typically $\sim$1/20) of their mass \cite{Bertone:2004pz}. The remaining dark matter particles would then populate the universe with a relic density which is of the same order as dark matter on cosmological scales \cite{Lee:1977ua,Krauss:1983ik,Kolb:1985nn,Scherrer:1985zt}. This ``WIMP miracle'' \cite{Steigman:1979kw} warrants extensive investigation, due to the possibility of observing the particle debris stemming from the occasional dark-matter pair-annihilation event in today's cold universe.

The probability of two dark matter particles ($\chi$) pair-annihilating into observable standard model particles is proportional to the thermally averaged pair-annihilation cross section times the relative particle velocity, $\langle\sigma v\rangle$, multiplied by the local particle dark matter number density squared, $n_\chi^2$. The first quantity, $\langle\sigma v\rangle$,  can be inferred from the requirement of having a relic abundance $\Omega_\chi\sim1/\langle\sigma v\rangle$ on the same order as the universal dark matter density, i.e. $\Omega_{\rm DM}\sim0.24$ \cite{Komatsu:2008hk}. The latter quantity is given by $n^2_\chi=(\rho_{\rm DM}/m_\chi)^2$, where $m_\chi$ indicates the mass of the dark matter particle $\chi$. Most dark matter annihilation processes are therefore predicted to occur in regions with a large dark matter density. Both intuition and detailed results from N-body simulations (see e.g. \cite{Diemand:2006ik, Diemand:2008in,Springel:2008by}) indicate that the center of the Milky Way galaxy is likely the brightest local dark matter annihilation site (barring the possibility of highly concentrated local dark matter clumps \cite{Brun:2009aj}). Among the possible signatures of dark matter annihilation, extensive studies have focused on gamma rays (see e.g. \cite{Baltz:2008wd,Jeltema:2008hf} and references therein) and neutrinos (for a recent study see \cite{Spolyar:2009kx}), particle species which have the benefit of carrying directional and spectral information. 

Other stable standard model particles produced in the pair annihilation of dark matter include charged cosmic rays such as electrons and positrons ($e^\pm$) as well as (anti-)protons and (anti-)deuterons. These charged species scatter off of magnetic field irregularities in the Galaxy, losing energy and diffusing before reaching the Earth. The random-walk propagation of charged cosmic rays in the Galaxy is usually described with a diffusion-loss equation and solved numerically \cite{Strong:1998pw} or semi-analytically \cite{Baltz:1998xv,Regis:2009qt}. A pedagogical review of cosmic-ray propagation and interactions in the Galaxy is given in Ref.~\cite{Strong:2007nh}.

The possibility of detecting an anomalous spectral feature in the flux of leptons, which could be traced back to the pair-annihilation of particle dark matter, has been long discussed (for early studies see e.g. Ref.~\cite{Rudaz:1987ry,Ellis:1988qp,Kamionkowski:1990ty,Profumo:2004ty}). This scenario has recently gained great momentum after results from the Pamela space-based antimatter detector reported an excess of high-energy (10-100 GeV) positrons over the assumed background of secondary positron production by inelastic cosmic-ray interactions \cite{Adriani:2008zr}. Tantalizingly, an excess (namely an anomalous ``bump'') in the flux of electrons plus positrons was also recently reported by the balloon-borne experiments ATIC \cite{2008Natur.456..362C} and PPB-BETS \cite{Torii:2008xu} at energies in the range of several hundred GeV. This excess was subsequently not confirmed by data from the Fermi Large Area Telescope (LAT), which, however, did indicate a much harder spectrum for the $e^\pm$ flux at high energy than previously assumed \cite{Abdo:2009zk,Grasso:2009ma}. This implies that the positron deficit reported by Pamela is at an even greater contrast with the standard expectation from cosmic ray models. The Fermi results agree with the low-energy range of other determinations of the $e^\pm$ flux \cite{Collaboration:2008aa}.

Although astrophysical sources such as pulsars \cite{Hooper:2008kg,Yuksel:2008rf,Profumo:2008ms,Malyshev:2009tw,Grasso:2009ma,Gendelev:2010fd} and supernova remnants \cite{Shaviv:2009bu,Blasi:2009hv} have been shown to provide a possible explanation for the positron fraction anomaly, dark matter models have been formulated which can fit both cosmic ray data and account for the positron excess (for a list of references see e.g. \cite{Profumo:2008ms}). In general, dark matter models that account for the Pamela anomaly need to have a large pair-annihilation cross section compared to the estimates from the standard thermal-relic calculation.  Furthermore, a mechanism or conservation law must be invoked to ensure that excess antiprotons are not produced at a detectable level, as the Pamela data place stringent constraints on any antiproton excess \cite{Adriani:2008zq}. Dark matter models which match both the Pamela and Fermi-LAT data commonly feature pair-annihilation modes dominated by positron production, and have rather large masses ($\sim$~1~TeV). Recently, Ref.~\cite{Kane:2009if} also pointed out that a wino-like candidate with a somewhat lighter mass can also account for the noted cosmic ray anomalies. We will entertain this possibility for one of the particle dark matter models considered below (although recent data on gamma-ray observations from local dwarf galaxies put stringent constraints on this category of models \cite{Collaboration:2010ex}). Winos dominantly pair-annihilate into $W^+W^-$ pairs, with a significant subsequent production of energetic $e^\pm$.

While the source term for the $e^\pm$ injected by dark matter pair annihilation traces the dark matter density profile squared, the resulting non-thermal population of electrons and positrons settles to an equilibrium configuration only after losing energy and spatially diffusing through scattering off of both the regular and turbulent components of the galactic magnetic fields. At high energy, the dominant energy loss processes are synchrotron and inverse Compton scattering, which create secondary radiation at radio frequencies as well as X-ray to gamma-ray energies, respectively (for a recent review of multi-wavelength emissions from dark matter annihilation see Ref.~\cite{Profumo:2010ya}). This secondary emission is a common denominator to any WIMP model. In particular, dark matter models that also explain the Pamela positron excess generically give very large signals, which are typically barely compatible with the extragalactic gamma-ray background, as shown e.g. in \cite{Profumo:2009uf,Huetsi:2009ex,Belikov:2009cx}. Constraints on dark matter models from the secondary radio and X-ray emission from $e^\pm$ produced in dark matter annihilation in extragalactic objects such as dwarfs and galaxy clusters has been studied in a number of recent analyses, see e.g. \cite{Totani:2004gy,Colafrancesco:2005ji,Baltz:2006sv,Colafrancesco:2006he,Profumo:2008fy,PerezTorres:2008ug,Jeltema:2008ax}.

\subsection{The role of synchrotron emission}

WIMP annihilation into charged leptons should produce synchrotron emission throughout a large spherical halo around the Milky Way. Placing the injection energy on the weak scale should create synchrotron emission at radio frequencies, since the galactic magnetic fields are of $\mathcal{O}\sim1-10\mu$G. This fact was envisioned long ago, see e.g \cite{Gondolo:2000pn,Bertone:2001jv,Aloisio:2004hy,Bergstrom:2006ny}, and, in conjunction with available radio data, was recently used to put severe constraints on the detectability of a gamma-ray signal from the galactic center \cite{Regis:2008ij} as well as on the viability of dark matter models that could explain the Pamela excess \cite{Bergstrom:2008ag}.

Interestingly, Finkbeiner pointed out in 2004 that a residual, unaccounted-for radio ``haze'' could actually be present in the data reported by the Wilkinson Microwave Anisotropy Probe (WMAP) \cite{Finkbeiner:2004us}, and that this radio emission could in principle be explained precisely with the radio haze predicted in typical WIMP models.\footnote{Throughout this paper we refer to the observed WMAP signal as the ``WMAP Haze" and our simulated results as the ``dark matter haze" or ``synchrotron haze"} This point was further elaborated upon in Ref.~\cite{Hooper:2007gi,Hooper:2007kb,Cholis:2008vb,Bottino:2008sv,Caceres:2008dr,Cumberbatch:2009ji, McQuinn:2010ju}, which essentially confirmed that the morphology and intensity of the haze from WIMP annihilation might match the observed radio excess in the WMAP data. Other studies have shown that pulsars can create an additional lepton component which adequately traces the morphology of the WMAP haze \cite{Kaplinghat:2009ix}. We note, however, that recent analyses by the WMAP team finds no evidence for a haze emission from the galactic center region in the polarization data \cite{Gold:2010fm}. Furthermore, the extraction of a haze may depend sensitively on the assumption that the 408 Mhz synchrotron map is morphologically similar to skymaps in the WMAP band \citep{Mertsch:2010ga}. Due to both the anomalous positron results, and in view of the recent successful launch of the Planck satellite \cite{Bouchet:2009tr}, there is considerable interest in models with a large annihilation rate into leptons. Thus, we consider this a timely moment to critically re-assess the diffuse radio emission from WIMP annihilation in the central regions of the Galaxy and to outline possible diagnostics that will allow to disentangle it from other possible astrophysical sources.

The secondary radio emission from dark matter annihilation depends on both the injection spectrum and the pair-annihilation rate of dark matter - quantities which are set by specifying the particle dark matter model at hand. In addition, the intensity and spatial distribution of the WIMP synchrotron halo depend on three astrophysical quantities: (1) the dark matter density distribution, (2) the diffusion setup, and (3) the magnetic field setup. While information on (1) can be gained from N-body simulations, with educated estimates for the role of baryons and stars in shaping the dark matter density distribution, a self-consistent guess for (2) and (3) is a difficult task. Thus, the production of accurate predictions involves not only factoring in as many astronomical and cosmic-ray data as possible, but also treating with accuracy the distribution of astrophysical cosmic-ray sources and the diffusion and energy losses of galactic cosmic rays.

The importance of the diffusion setup, including in particular the geometry of the diffusion zone, the magnitude of the diffusion coefficient and its rigidity dependence, as well as the spatial distribution of Galactic magnetic fields, cannot be understated in the context of calculating the radio emission from dark matter annihilation. For instance, the height of the diffusion zone determines at which galactic latitude the $e^\pm$ population responsible for the radio emission leaks out and is artificially truncated by the boundary conditions imposed to the diffusion-loss equation; similarly, the spatial distribution of magnetic fields determines the intensity of the synchrotron emission, which scales quadratically with the magnitude of the magnetic fields. 

In the present work we study the morphology of the synchrotron halo produced by electrons and positrons created via dark matter annihilation in the context of {\em self-consistent diffusion models}. In particular, we interface the most complete publicly available cosmic-ray propagation code, {\tt Galprop} \cite{Strong:1998pw}, with the spectra $e^\pm$ from dark matter annihilation calculated by the {\tt DarkSUSY} package \cite{Gondolo:2004sc}. Testing various WIMP annihilation pathways, we compare our observations to constraints from cosmic-ray observations (e.g. Boron to Carbon and $^{10}$Be to $^9$Be). Imposing this set of observational constraints on the diffusion setup forces the morphology of the dark matter ``haze'' to specific configurations, which we outline in detail and compare to the WMAP data residual as analyzed in Ref.~\cite{Dobler:2007wv}. 

Finally, we explore the diffuse inverse Compton emission from the up-scattering of the inter-stellar radiation field, which stems from the same population of dark-matter-induced $e^\pm$ that would produce the radio haze. We then compare the predicted emission with the Fermi diffuse gamma-ray data, and note the ability for observations of $\gamma$-ray emission to put strong constraints on models of the WMAP haze.

The outline of this study is as follows: in the next section we describe the ``baseline'' setup adopted for the benchmark particle dark matter models  (Sec.~\ref{sec:dmmodels}) and for the set of galactic dark matter density profiles (Sec.~\ref{sec:dmdensity}). In addition, we provide details on the diffusion setup for cosmic ray propagation we employ in this analysis (Sec.~\ref{sec:diffusion}). We study how the morphology of the synchrotron emission from dark matter annihilation depends on the diffusion setup in Sec.~\ref{sec:morpho}, on the dark matter density profile in Sec.~\ref{sec:morphodm}, and on the spatial distribution of magnetic fields in Sec.~\ref{sec:morphob}. We then study the impact of matching the adopted diffusion scenarios to cosmic ray data in Sec.~\ref{sec:cosmicrays}, and outline the expected spectrum of radio emission and inverse Compton emission in the context of the self-consistent scenarios in Sec.~\ref{sec:radio} and~\ref{sec:invco}, respectively. Finally, we give an overview and a discussion of our results and predictions in Sec.~\ref{sec:disc}.

%%%%%%%%%%%%%%%%%%%%%%%%%%%%%%%%%%%%%%%%%%%%%%%%%%%%%%%%%%%%%%%%%%%%

\section{Baseline Setup}
In this section we define a baseline setup for the present study. In particular, we discuss the benchmark particle dark matter models we employ (sec.~\ref{sec:dmmodels}), and the set of Galactic dark matter density distribution profiles (sec.~\ref{sec:dmdensity}). Both the particle setups and the density distributions are chosen with the intention of illustrating extreme cases, and of capturing the full range of possible outcomes for our subsequent calculations. We then discuss our treatment of diffusion and energy losses, specifying the details of the setup we employ here as well as the assumptions on the magnetic field spatial distribution in sec.~\ref{sec:diffusion}.

\subsection{Dark Matter Particle Models}\label{sec:dmmodels}

The injection spectrum of electrons and positrons from particle dark matter annihilation depends on (a) the dark matter mass, which sets the energy scale for the spectrum, and (b) the relative probability for the dark matter to pair-annihilate into given Standard Model final states. These branching ratios determine the spectral shape of the injected $e^\pm$, which is calculated from the spectral distribution of electrons and positrons resulting from a given dark matter particle state. These spectral distributions are calculated by interpolating results from Monte Carlo simulations, a task which is carried out automatically in the {\tt DarkSUSY} package \cite{Gondolo:2004sc}. Finally, the overall normalization is directly proportional to the thermally averaged pair annihilation cross section times the relative velocity, $\langle\sigma v\rangle$, for $v\to0$.

Electrons and positrons are produced in dark matter annihilation events with spectral features which depend on the production mechanism. In particular, they can be produced (1) promptly, (and with an energy $E_{e^\pm}\simeq m_\chi$) in the process $\chi\chi\to e^+e^-$, (2) with a {\em hard} spectrum, if they originate from muon or tau lepton decays, or (3) with a {\em soft} spectrum if they stem from the decay of charged pions resulting from hadronization chains of strongly interacting particles or from the hadronic decay modes of non-strongly interacting particles (tau leptons, gauge and Higgs bosons). Several theoretically motivated dark matter particle setups predict dominant pair-annihilation into a pair of charged SU(2) gauge bosons (this is the case, for instance, in anomaly mediated supersymmetry breaking \cite{Feng:1999fu}). In this case, the predicted $e^\pm$ injection spectrum is a combination of the above mentioned soft and hard spectral shapes. The soft channels correspond to the hadronic decay modes of the $W$, and the hard channels to the $W\to l \nu_l$ leptonic modes, which produce energetic $e^\pm$ either promptly or via lepton decay.

In the present study, we try to encompass the largest possible variety of WIMP dark matter $e^\pm$ injection spectra, with a preference toward models that are theoretically and/or phenomenologically motivated. For consistency, we set the global cross-section for all dark matter particles to be $\langle\sigma v\rangle=3\times 10^{-26}\ {\rm cm}^3/{\rm s}$ and evaluate the necessary boosts from this value in order for our models to match the intensity of the WMAP haze. We note that this implies large boost factors for the case of very massive dark matter particles to get a significant signal.

Specifically, we consider a ``vanilla'' dark matter model which corresponds to realizations of minimal supersymmetric extensions of the Standard Model as a benchmark for a soft $e^\pm$ injection spectrum. To populate the halo with relatively low-energy $e^\pm$ from WIMP annihilation, we set the dark matter mass to 40 GeV. This corresponds to the lightest bino-like neutralino in models with gaugino unification at the grand unification scale. It is customary to assume, for light enough dark matter particles, a $b\bar b$ final state, given that in the case of supersymmetry, if both the $t\bar t$ channel is kinematically closed and decays into gauge or Higgs bosons are suppressed by the neutralino composition, helicity suppression dictates that the dominant fermion-antifermion final state corresponds to the heaviest fermion. We dub this model as {\bf Soft}, from its spectral shape and mass range.

We choose our opposite extremum benchmark to have a hard $e^\pm$ spectrum and a large mass. Specifically, we select a model with exclusive pair-annihilation into muons, i.e. $\chi\chi\to\mu^+\mu^-$. These models are motivated by successful fits to the Pamela positron anomaly data as well as to the Fermi-LAT $e^\pm$ data, as shown for instance in Ref.~\cite{Bergstrom:2009fa}. A further requirement for this type of phenomenologically motivated setup is a rather large mass scale, which we fix here at 1.5 TeV. Theoretical frameworks where particle models with similar $e^\pm$ injection spectra, and comparably large pair annihilation cross sections were proposed for instance in Ref.~\cite{ArkaniHamed:2008qn} and \cite{Nomura:2008ru}. We indicate this model as {\bf Hard} because of its spectral appearance.

%%%%%%%%%%%%%%%%%%%%%%%%%%%%%%%%%%%%%%%%%%%%%%%%%%%
\begin{table}[!b]
\begin{tabular}{|c |c |c |c|}
\hline
Model ID & Mass (GeV) & $\langle\sigma v\rangle/({\rm cm}^3/{\rm s})$ & Ann. Final State \\
\hline
{\bf Soft} & 40 & $3\times10^{-26}$ & $b\bar b$\\
{\bf Hard}& 1500 & $3\times10^{-26}$ & $\mu^+\mu^-$\\
{\bf Wino}& 200 & $3\times10^{-26}$ & $W^+W^-$\\
\hline
\end{tabular}
\caption{\label{tab:models}\it\small Physical properties of the benchmark particle dark matter models we consider in this study.}
\end{table}
%%%%%%%%%%%%%%%%%%%%%%%%%%%%%%%%%%%%%%%%%%%%%%%%%%%

Finally, to interpolate between the mass ranges and the spectral shapes, and to pick an alternative particle dark matter setup that explains the anomalous Pamela positron excess (see however the recent results of Ref.~\cite{Collaboration:2010ex}), we consider a model along the lines of those discussed in Ref.~\cite{Kane:2009if}. Specifically, we consider a mass of 200 GeV and model a wino-like neutralino dark matter as a particle with  a 100\% pair-annihilation branching ratio into $W^+W^-$ pairs. We dub this model as {\bf Wino}. Notice that we do not assume a pair-annihilation cross section at the level expected for wino-like dark matter \cite{Kane:2009if}, in order to compare directly the boost factors among the three models under consideration. The ``boost factor'' corresponding to a 200 GeV wino would be on the order of $10^2$.

%%%%%%%%%%%%%%%%%%%%%%%%%%%%%%%%%%%%%%%%%%%%%%%%%%%
\begin{figure*}[!t]
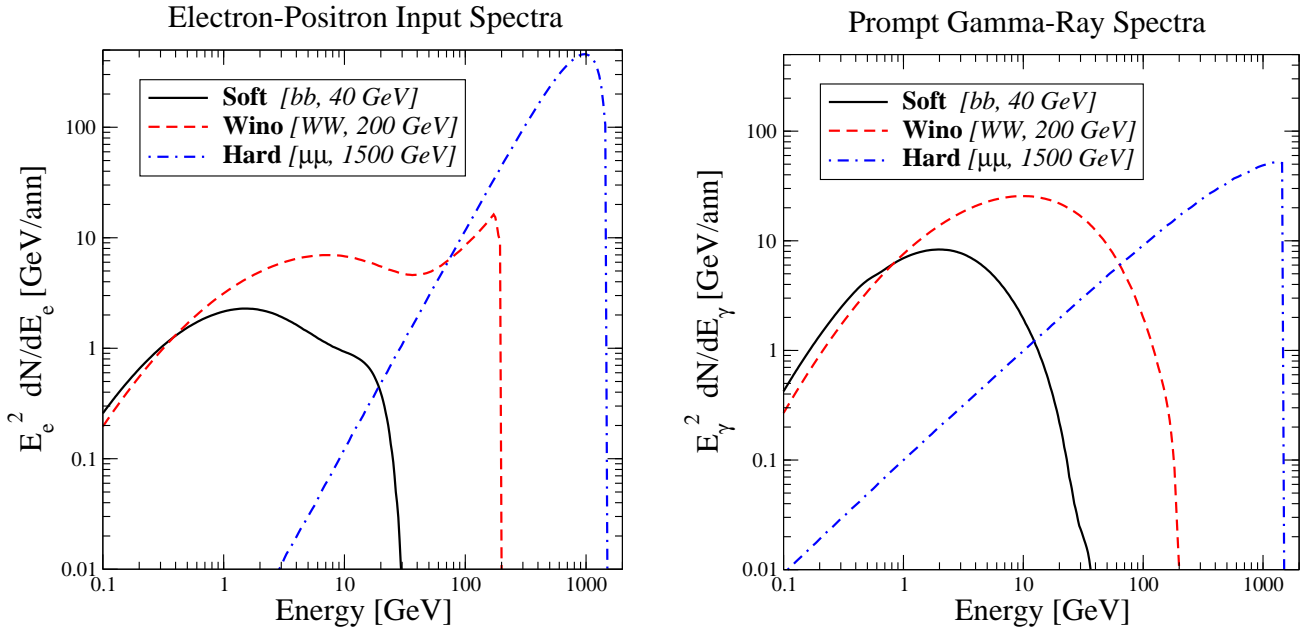

\mbox{\includegraphics[width=0.5\textwidth,clip]{f1a.eps}\quad\quad\includegraphics[width=0.5\textwidth,clip]{f1b.eps}}
\caption{\label{fig:input}\it\small Left: The spectrum (multiplied by energy squared) of electrons (and positrons) produced by the pair annihilation of dark matter for our three benchmark models: {\bf Soft} (featuring a lightest neutralino mass of $m_\chi=40$ GeV and a dominant annihilation final state into $\bar b b$), {\bf Wino} ($m_\chi=200$ GeV, $W^+W^-$) and {\bf Hard} ($m_\chi=1.5$ TeV, $\mu^+\mu^-$). Right: the same, but for the spectral energy distribution of gamma rays promptly produced at dark matter annihilation, for the three benchmark models.}
\end{figure*}
%%%%%%%%%%%%%%%%%%%%%%%%%%%%%%%%%%%%%%%%%%%%%%%%%%%

In Tab.~\ref{tab:models} we summarize the physical properties of the benchmark dark matter models we employ in the present study: namely the particle mass, its thermally averaged pair-annihilation cross section times velocity and its dominant annihilation final state. Fig.~\ref{fig:input}, left, shows the differential $e^\pm$ injection spectrum (prior to propagation and energy losses) for the three models under consideration, where the flux is multiplied by energy squared and in units of GeV per annihilation. The spectral properties discussed above are clearly visible in the plot, including: the soft $e^\pm$ spectrum from pion decay in the case of the $b\bar b$ final state, the mixed soft-hard spectrum of winos, with a peak corresponding to the leptonic production modes, and the $\sim E^{-1}_{e^\pm}$ spectrum from $\mu^+\mu^-$ decay. 

We also show in the right panel of Fig.~\ref{fig:input} the prompt gamma-ray spectrum produced in the same annihilation event that produces the $e^\pm$. While the $b \bar b$ and the $W^+W^-$ spectra are dominated by the two-photon decay of neutral pions, the spectrum of gamma rays from the muon final state entirely stems from final state radiation, and scales as $E^{-1}_\gamma$ (see e.g. \cite{Beacom:2004pe}). This ``primary'' emission will also play an important role when estimating the overall gamma-ray emission from dark matter annihilation in the central regions of the Galaxy; although the spectral details and the normalization of this component are not directly related to the secondary emission which is the core of the present study.

\subsection{Dark Matter Galactic Density Profiles}\label{sec:dmdensity}

The intensity of indirect dark matter signals from both prompt and secondary emission (inverse Compton and bremsstrahlung at high energy, synchrotron at low energy) are highly dependent on the distribution of dark matter density throughout the Galaxy. In the case of annihilating dark matter, the total interaction rate goes as the integral over the local dark matter density squared. We note that in the case of decaying dark matter, the annihilation rate goes instead as the integral over the local density. 

In this study we restrict our evaluation to several spherically symmetric dark matter profiles. We fix the local dark matter density at the solar position ($R_\odot\simeq 8.5$ kpc) to be equal to $\rho(R_\odot)=0.389\ {\rm GeV}/{\rm cm}^3$ \cite{2009arXiv0907.0018C}. We express each of the dark matter density profiles as functions $f(x)$ of the dimensionless variable $x=R/a$, where $R$ is the distance from the center of the Galaxy and $a$ is a scale radius for the density profile.

Motivated by recent high resolution collisionless N-body simulations \cite{Pieri:2009je}, we adopt the best-fitting Einasto density profile obtained from the results of the {\em Aquarius} simulation \cite{2008MNRAS.391.1685S}: 

\begin{equation}
f_{\rm Einasto}(x)=\exp\Big[-\frac{2}{\alpha}\left(x^\alpha-1\right)\Big],\quad\alpha\simeq0.17
\end{equation}
and a profile extrapolated from the {\em Via Lactea II} simulation \cite{2008MNRAS.391.1685S, Diemand:2008in}:
\begin{equation}
f_{\rm Via\ Lactea}(x)=x^{-\gamma}(x+1)^{-3+\gamma} \quad \gamma\simeq1.24.
\end{equation}
The scale radius for $f_{\rm Einasto}$ is assumed to be $a_{\rm Aquarius}\simeq 22$ kpc, a value which also matches what found in Ref.~\cite{2008MNRAS.391.1685S}, while that for $f_{\rm Via\ Lactea}$ is set to $a_{\rm Via\ Lactea}\simeq 28.1$ kpc \cite{2008MNRAS.391.1685S}.

Ref.~\cite{2008MNRAS.391.1685S} points out that their results can be fitted with the prototypical Navarro-Frenk-White density profile \cite{Navarro:1996gj}, for which:
\begin{equation}
f_{\rm NFW}(x)=x^{-1}(x+1)^{-2}
\end{equation}
and where we assume $a_{\rm NFW}\simeq22$ kpc. To illustrate the effect of the scale radius, we will also consider values for $a_{\rm NFW}$ as large as 100 kpc and as small as 10 kpc.

Physical processes related to the accretion of baryons clumped within the central regions of the Galaxy \cite{ElZant:2001xb,ElZant:2001re} can lead to a smoothing of the central cusps of the profile obtained in collisionless dark matter simulations, and to the formation of large cores. The resulting density profiles can be modeled with profiles similar to those proposed in Ref.~\cite{1995ApJ...447L..25B,2000ApJ...537L...9S}, which we indicate as a Burkert profile, with a functional form:

\begin{equation}
f_{\rm Burkert}(x)=(1+x)^{-1}(1+x^2)^{-1}
\end{equation}

Galactic dynamics data point to $a_{\rm Burkert}\simeq 11.6$ kpc. For illustrative purposes, we will consider a range of scale-sizes, with $a_{\rm Burkert}\simeq 5$~kpc, $a_{\rm Burkert}\simeq 20$~kpc, and $a_{\rm Burkert}\simeq 50$~kpc.

Several models of the galactic dark matter distribution predict cusps of high dark matter density near the galactic center. In this study, we do not include such cusps. In Section V, we shall point out that the impact of dark matter cusps on the morphology of the dark matter synchrotron haze can be understood and extrapolated off of our non-cusped models.

\subsection{Diffusion Models}\label{sec:diffusion}
We utilize the state-of-the-art cosmic ray propagation code {\tt Galprop} \citep{Strong:1998pw, 2009arXiv0907.0559S}, which was created to solve cosmic-ray transport in a self-consistent manner. {\tt Galprop} contains modules to compute secondary to primary ratios, synchrotron radiation, inverse Compton scattering, $\pi^{0}$-decay, and bremsstrahlung radiation, from the same astrophysical inputs. These calculations include effects such as nuclear spallation and energy loss due to interactions between cosmic rays and the Galaxy's gas, radiation, and magnetic fields. 

Utilizing a 2D simulation, we choose a standard diffusion model (GALDEF 50p\_599278)\footnote{http://galprop.stanford.edu/GALDEF/galdef\_50p\_599278}, based on the latest public version of the {\tt Galprop} code. We note that the parameter space for propagation of cosmic rays is very large, and thus we take default assumptions consistent with the above model unless otherwise noted. In Table~\ref{standard_diffusion}, we show the default values for the parameters most relevant to our study. We will evaluate first order changes in each of these parameters, while keeping the remaining terms constant to the values indicated in the Table. We place upper and lower bounds on each parameter, the values of which are not necessarily physically motivated. Instead, these parameters will serve to illustrate the effect of each individual parameter on the diffusion of high energy lepton sources. In Sec.~\ref{sec:cosmicrays}, we will examine examples of the interdependence between diffusion parameters, as well as limiting constraints on possible model parameters. In order to model the galactic ICS emission from our lepton population (see Section IX), we employ the default Galprop model for the interstellar radiation field (ISRF) \citep{Porter:2005qx}.

We note that the assumed galactic diffusion setup may not be constant throughout the galaxy. Specifically, diffusion may be significantly different near the galactic center, due, for instance, to high gas and dust densities, which may affect the morphology of both synchrotron signals~\citep{Biermann:2009td} as well as local cosmic ray abundances~\citep{Gebauer:2009hk}. In Section VII we discuss the effect of these parameters on our simulated synchrotron morphologies.

In order to determine the e$^\pm$ fluxes from each of our dark matter annihilation pathways, we have modified the {\tt Galprop} code to accept the input $e^\pm$ spectrum directly from the output of the {\tt DarkSUSY} code \citep{2004JCAP...07..008G}. The {\tt DarkSUSY} code has been created to calculate the decay branching ratios and energy spectra for neutralino dark matter models as an input to the determination of dark matter direct and indirect detection rates. {\tt DarkSUSY} is still the state-of-the-art tool in calculating the injection spectra of stable Standard Model particles which result from given particle dark matter setups. The {\tt Galprop} code then interpolates between the input injection energy spectrum of e$^\pm$, and calculates the approximate resulting flux for each necessary lepton energy.

%%%%%%%%%%%%%%%%%%%%%%%%%%%%%%%%%%%%%%%%%%%%%%%%%%%
\begin{table}
\begin{tabular}{l | l | l | l | l}
\hline\hline
\label{standard_diffusion}
Parameter & Default Value & Lower Bound & Upper Bound & Alternative\\
\hline
Radius & 20 kpc & --- & --- & --- \\
Height & 4 kpc & 2 kpc & 16 kpc & ---\\
D$_{0}$ & 5.0 x 10$^{28}$ cm$^2$s$^{-1}$ & 1.0 x 10$^{28}$ cm$^2$s$^{-1}$ & 1.0 x 10$^{30}$ cm$^2$s$^{-1}$ & --- \\
B$_{galaxy}$ & 11.69e$^{-r/10 {\rm kpc}}$e$^{-z/2 {\rm kpc}}\mu$G & 11.69e$^{-r/10 {\rm kpc}}$e$^{-z/1 {\rm kpc}}\mu$G & 10$\mu$ G & 11.69e$^{-r/10 {\rm kpc}}$e$^{-z/8 {\rm kpc}}\mu$G \\
v$_\alpha$ & 25 km s$^{-1}$ & 0 km s$^{-1}$ & 100 km s$^{-1}$ & --- \\
Convection & Disabled & --- & 100 km s$^{-1}$ & 50 km s$^{-1}$\\
\hline\hline
\end{tabular}
\caption{\it List of important {\tt Galprop} parameters used throughout this paper. Our default values are listed in the 2nd column, while important parameter space checks employed throughout this paper are listed in subsequent columns.}
\end{table}
%%%%%%%%%%%%%%%%%%%%%%%%%%%%%%%%%%%%%%%%%%%%%%%%%%%

\section{The Haze Morphology: Default Model}

From our {\tt Galprop} simulations, we calculate the spatial distribution of synchrotron radiation over the line of sight from the solar position. In order to determine the intensity of synchrotron radiation which stems from products of dark matter annihilation, we create simulations both with and without dark matter for each set of diffusion parameters. Subtracting the dark matter signal, pixel by pixel, from the full simulations leaves a residual which stems purely from dark matter annihilations.

\begin{figure*}
\mbox{\includegraphics[width=0.5\textwidth,clip]{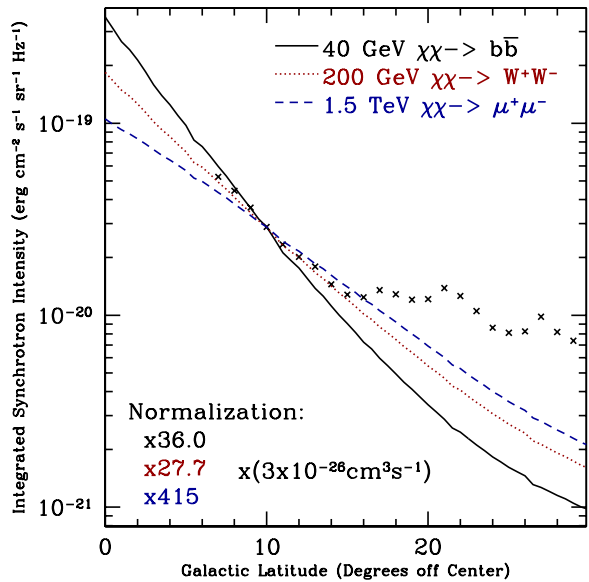}}
\caption{\label{baseplot}\it\small Simulated dark matter haze for three dark matter channels (40 GeV $\chi\chi$ -$>$ $b\bar{b}$ ({\bf Soft}, black solid), 200 GeV $\chi\chi$ -$>$ W$^{+}$W$^{-}$ ({\bf Wino}, red dotted), 1500 GeV $\chi\chi$ -$>$ $\mu^{+}\mu^{-}$ ({\bf Hard}, blue dashed)) plotted against the WMAP haze as determined by \citet{Dobler:2007wv}.}
\end{figure*}

For the observed WMAP haze, we use the fits derived from CMB5 in \citet{Dobler:2007wv}. However, we fit the foregrounds over the entire sky, so that the analog of Eq. (18) in \citet{Dobler:2007wv} becomes:

\begin{equation}
{\bf r}_H = r_{FS8} + a_hH
\end{equation}

The resulting map is then mean-subtracted to remove any unwanted residuals stemming from inadequate subtraction of the extragalactic isotropic background, and the data is displayed in unmasked regions (corresponding to the \citet{Dobler:2007wv} mask) averaging the longitudinal values over the angular window -10$^\circ$~$<$~l~$<$~10$^\circ$, in order to decrease any spurious noise effect.

In Figure~\ref{baseplot}, we show the synchrotron overabundance at 23 Ghz attributable to dark matter in our default simulation, compared against observations of the WMAP haze. We multiply the output from our dark matter model by a fixed constant in order to normalize the diffuse radio emission (``haze'') to the WMAP haze at a position of 10$^\circ$ latitude. This normalization accounts non-discriminantly for any change in the dark matter annihilation rate, but does not model the spatial characteristics which may be attributable to e.g. Sommerfeld enhancement or dark matter substructure. Since the boost factor is a well known uncertainty in calculating dark matter annihilation rates, we only concern ourselves here with matching the latitudinal distribution of dark matter, which depends primarily on the much better constrained global distribution of dark matter.

In examining Figure~\ref{baseplot}, we note a significant contrast between the spatial distribution of the observed WMAP haze and our default simulation parameters. While the WMAP haze falls off by approximately a factor of 6.7 between 7$^\circ$ and 30$^\circ$ of latitude, our default parameters show a much steeper trend, falling off by factors of $\sim$61 for our {\bf Soft} model, $\sim$31 for our {\bf Wino} model, and $\sim$21 for our {\bf Hard} model. Because this discrepancy does not depend on the overall annihilation rate, it is difficult to explain with variations in the dark matter annihilation setup. Instead, this discrepancy would most likely be explained by astrophysical parameters, and thus we first examine the role of our chosen diffusion parameters in determining the spatial distribution of the dark matter haze. 

\section{The Haze Morphology: Role of Diffusion Parameters}\label{sec:morpho}

\begin{figure*}
\mbox{\includegraphics[width=0.5\textwidth,clip]{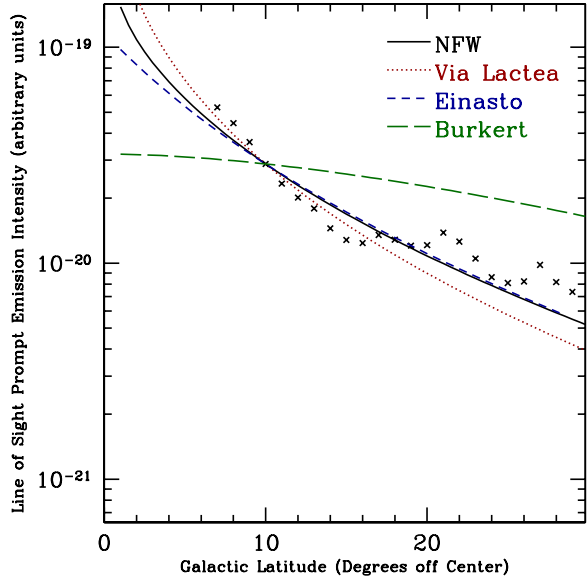}}
\caption{\label{nodiffusion}\it\small Simulated dark matter haze for four dark matter density profiles: NFW (black line), Via Lactea (red dotted), Einasto (blue dashed), and Burkert (green long dashed) for models in which no diffusion occurs and the secondary electrons stay exactly at the point of annihilation. The parameters for each dark matter profile are given in Sec.~\ref{sec:dmdensity}.}
\end{figure*}

In order to illustrate the importance of our diffusion setup in determining the morphology of the WMAP haze, we show in Figure~\ref{nodiffusion} the simulated WMAP haze for models in which we have entirely disabled diffusion (setting the diffusion constant,  Alfv\'en and convection velocity to 0). In this model, the dark matter synchrotron overabundance is a tracer of the position of dark matter annihilation and is identical to both the integral over the line of sight of the dark matter density squared, and the integral over the line of sight of prompt emission from dark matter annihilation. Interestingly, we note that in this case, the ratio between 7$^\circ$ and 30$^\circ$ is actually closer to unity than in our default simulation. This suggests that cosmic ray diffusion does not help equilibrate medium and high latitude emission, but instead greatly increases secondary emission near 10$^\circ$ by moving a large number of electrons from dark matter near the galactic center up to small galactic latitudes. 

\begin{figure*}
\mbox{\includegraphics[width=1.0\textwidth,clip]{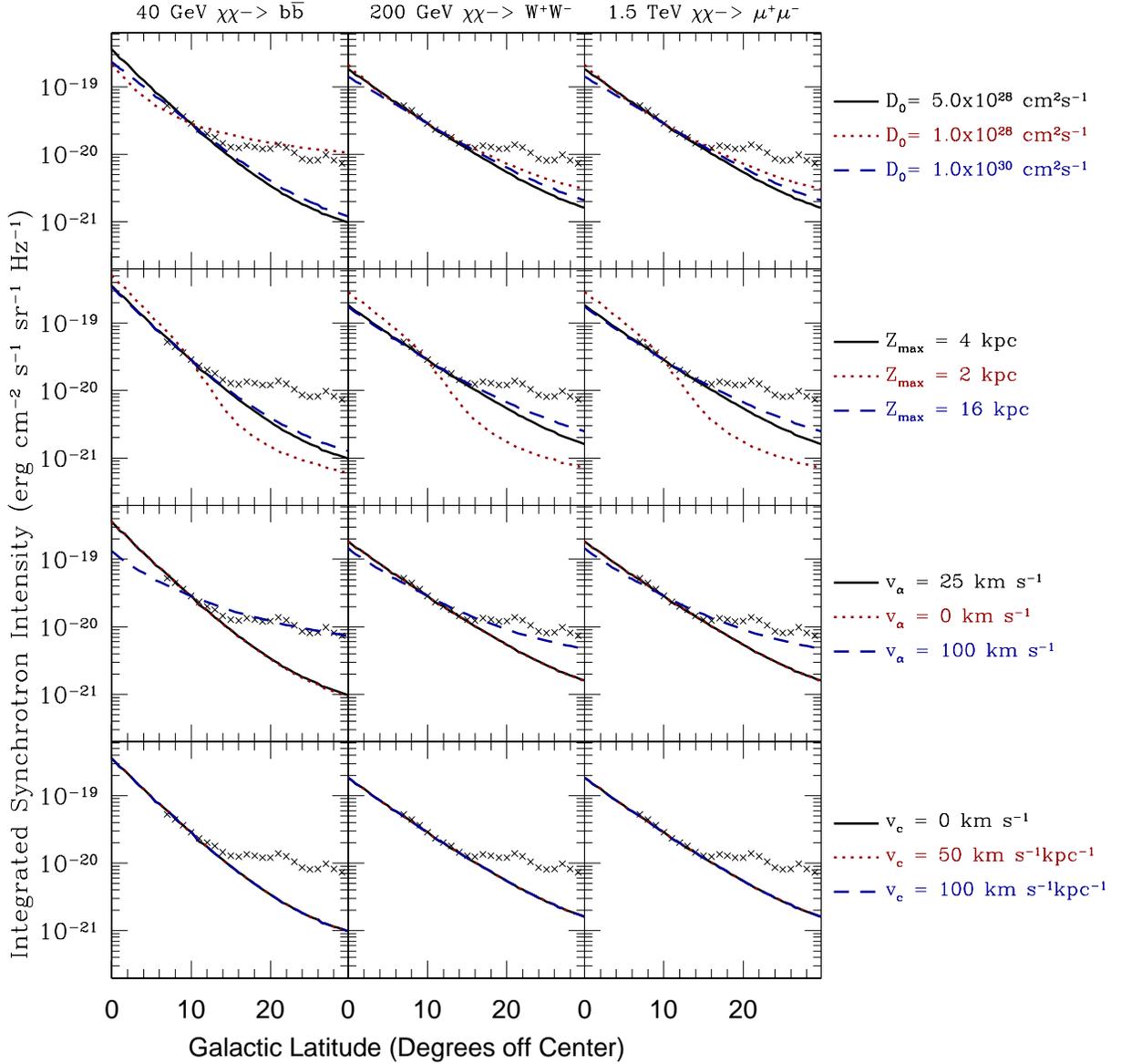}}
\caption{\label{diffusionplot}\it\small Simulated dark matter haze for three dark matter channels (40 GeV $\chi\chi$ -$>$ $b\bar{b}$ ({\bf Soft}, left), 200 GeV $\chi\chi$ -$>$ W$^{+}$W$^{-}$ ({\bf Wino}, center), 1500 GeV $\chi\chi$ -$>$ $\mu^{+}\mu^{-}$ ({\bf Hard}, right)) for illustrative choices of four important diffusion parameters (diffusion constant (top), diffusion region height (2nd row), Alfv\'en velocity (3rd row), and convection velocity (bottom) plotted against the WMAP haze as determined by \citet{Dobler:2007wv}. In each plot, our default simulation is shown in solid black, with changes in each parameter shown in dotted red and dashed blue.}
\end{figure*}

Since changes in the diffusion setup can produce counter-intuitive changes like those shown above, it is important to create models of the dark matter haze over wide ranges of diffusion setups, in order to determine whether any of these provides a reasonable match to the WMAP haze. To determine the first order effects of changes in the {\tt Galprop} diffusion setup on the spatial distribution of the dark matter haze, we modulate each diffusion parameter independently around its default value while keeping the remaining diffusion parameters constant. In Figure~\ref{diffusionplot}, we plot the synchrotron overabundance attributable to dark matter at 23 Ghz for changes in the diffusion constant (D$_0$),  diffusion region height (Z$_{max}$), Alfv\'en velocity (v$_{\alpha}$), and convection velocity (v$_{conv}$) for each of our dark matter annihilation pathways. We again normalize each model to match the WMAP haze at 10$^\circ$ latitude. In each plot, the black solid line shows our default propagation setup, while the various colored lines show illustrative changes in a single propagation parameter.

As shown in Figure~\ref{nodiffusion}, cosmic ray diffusion actually increases the falloff in emission between 7$^\circ$ and 30$^\circ$. However, on the opposite end of the spectrum, we would expect completely isotropic emission  in the limit of infinitely large diffusion. Interestingly, we find that the standard diffusion constant of D$_0$~=~5.0~x~10$^{28}$~cm$^2$s$^{-1}$ lies near a maximum in the ratio of interior (i.e. low-latitude) synchrotron production to high latitude emissivity. Thus, moving away from this value of the diffusion constant in either direction helps bring around a better match to the WMAP haze morphology. We find the effect of limiting diffusion from the galactic center to low latitude regions to be dominant, as low values of the diffusion constant tend to greatly increase the ratio of high latitude to low latitude synchrotron radiation, providing a best match to the WMAP haze for our {\bf Soft} model at  D$_0$~$\approx$~1.0~x~10$^{28}$~cm$^2$s$^{-1}$ (consistent with the results of \citet{Hooper:2007kb}), while our {\bf Wino} and {\bf Hard} models require even lower diffusion constants. High values of the diffusion constant provide a slightly better match to the WMAP haze, but the effect is never strong enough to provide a reasonable match, even for extremely large diffusion constants (D$_0$~=~1.0~x~10$^{32}$~cm$^2$s$^{-1}$ or greater).

Changes in the height of the diffusion zone (2nd row) have an interesting effect on dark matter models of the haze. For smaller diffusion zones (Z$_{max}$~=~2~kpc), we see a rapid, and functionally different, decrease in the dark matter haze starting at approximately 10$^{\circ}$ latitude. This is due to the fact that from the solar position, a latitude of 10$^{\circ}$ implies a height of 1.47~kpc above the galactic center. Any observation at higher angles will quickly lose large quantities of $e^\pm$ emitting synchrotron radiation to the confines of the simulation zone. This restricts the normal method of making corresponding changes in the diffusion constant and simulation height (as is commonly done for studies of the local primary to secondary ratios). However, from the simulation for Z$_{max}$~=~16~kpc, we find that even large increases in the diffusion zone beyond our default value create only small changes in the output haze. Furthermore, increases in the height of the simulation region will require corresponding increases in the diffusion constant. Thus, we find that a diffusion height of 4~kpc is an optimal value for simulations of the WMAP haze up to 30$^{\circ}$ latitude.

The Alfv\'en velocity can affect the angular distribution of the synchrotron haze by altering the diffusion of dark matter secondaries in momentum space \citep{1990acr..book.....B, 1994ApJ...431..705S}. The diffusion through momentum space varies as the square of the Alfv\'en velocity:

\begin{equation}
D_{pp} \propto \frac{p^2V_\alpha^2}{9D_{xx}}
\end{equation}

and thus changes which decrease the Alfv\'en velocity have little effect on the angular distribution of the haze. However, large Alfv\'en velocities can create very effective diffusion through momentum space, which either propel particles up to high latitudes, or restrict particles from entering even the low latitude simulations regions. This mimics the improvements seen in diffusion constants that are both higher and lower than our standard setup. We find that an Alfv\'en velocity of approximately 100~km~s$^{-1}$ can successfully match the morphology of the WMAP haze for each of our dark matter particle physics models.

While diffusion is the dominant driving force for spatial diffusion of cosmic rays, galactic convection can also play a significant role. As we have convection disabled in our base model, we test the parameter space for models with  stronger convection. In the bottom row, we test several convection velocities which linearly vary with the height above the galactic plane. We note that changes in the convection velocity do not greatly change the spatial distribution of the dark matter haze, and the variation they do create goes in the wrong direction to match the observed haze morphology. This is due to the fact that convection is normally strongest at higher latitudes, and thus efficiently moves high energy leptons out of our simulation. We also note that the effect of convection is strongest for soft dark matter models, as the effect of convection is most important between 1-10 GeV, which is a sweet spot for e$^\pm$ production from our soft model.

\begin{figure*}
\mbox{\includegraphics[trim=203mm 203mm 200mm 70mm, clip, width=18.5cm, angle=180]{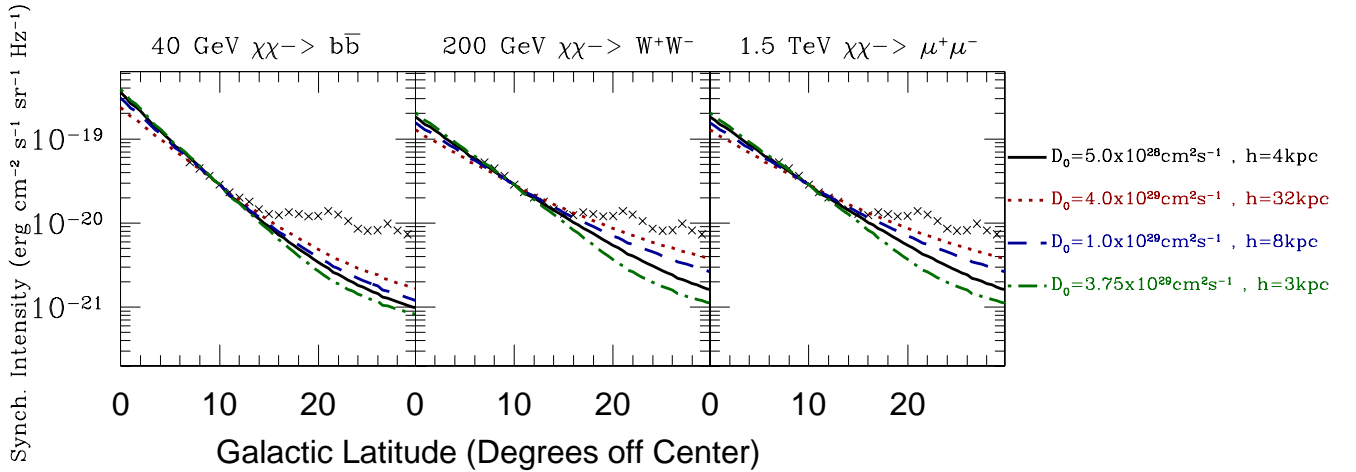}}
\caption{\label{dzhplot}\it\small Simulated dark matter haze for three dark matter channels (40 GeV $\chi\chi$ -$>$ $b\bar{b}$ ({\bf Soft}, left), 200 GeV $\chi\chi$ -$>$ W$^{+}$W$^{-}$ ({\bf Wino}, center), 1500 GeV $\chi\chi$ -$>$ $\mu^{+}\mu^{-}$ ({\bf Hard}, right)) for illustrative choices of the diffusion constant and simulation height which keep their ratio constant. In each plot, the black solid line shows our default propagation setup, while the various colored lines show our parameter space models.}
\end{figure*}

Lastly, we note that a well known degeneracy exists between changes in the diffusion constant and the scale height of the diffusion region. While a change in the direction towards low diffusion constants and low scale heights are unlikely to create a match to the WMAP haze, as scale heights less than 4~kpc cut off emission above the galactic center at latitudes near 30$^\circ$. However, in Figure~\ref{diffusionplot}, we see that increases in both the diffusion constant and the height of the diffusion region increase the intensity of high latitude synchrotron relative to emission at 10$^\circ$ below the galactic center. This raises the possibility that corresponding changes in these parameters may create better dark matter fits to the WMAP haze, while preserving the local cosmic ray density.

In Figure~\ref{dzhplot} we show the resulting dark matter haze for several parameter choices which leave the ratio of D$_0$/height constant. We find that larger diffusion constants and heights do increase the relative intensity of high latitude synchrotron radiation. However, we do not begin to see a good match until we employ extreme cases such as a diffusion height of (z=32 kpc), which itself still falls short of reproducing a match to the WMAP haze morphology, with reasonable matches occuring only in the cases of our harder spectra. We strongly caution the reader, however, about the accuracy of the synchrotron morphology for this extreme height, since the galactic magnetic field is almost completely unknown in this region. Nevertheless, as this is an important degeneracy condition in our work, we will examine it further in Section~\ref{sec:cosmicrays}.

\section{The Haze Morphology: Role of Dark Matter Density Profiles}\label{sec:morphodm}

The distribution of dark matter secondaries should depend strongly on the galactic distribution of dark matter annihilations. Throughout this paper, we model only the smooth component of dark matter, normalizing the result to the synchrotron haze in order to account for uncertainties in the dark matter annihilation rate. Since we are concerned with annihilating dark matter, we note that the number of annihilations at any point in space varies with the square of the local dark matter density. To test the effect of various dark matter density profiles on our simulated haze, we examine several common dark matter distributions in addition to our default NFW profile. We test the functional forms given by the Via Lactea simulation (R$_{sc}$~=~28.1~kpc) \citep{Diemand:2008in}, an Einasto profile (R$_{sc}$~=~20~kpc, $\alpha$~=~0.17) \citep{2008MNRAS.391.1685S}, and a Burkert profile (R$_{sc}$~=~11.6~kpc) \citep{1995ApJ...447L..25B,2000ApJ...537L...9S} . For all models we adopt a solar distance R$_\odot$~=~8.5~kpc and a local dark matter density of 0.389~GeV~cm$^{-3}$ \citep{2009arXiv0907.0018C}. In Figure~\ref{nodiffusion}, we have showed the normalized synchrotron intensity when the motion of dark matter secondaries is neglected. We note that this is identical (within a normalization constant) to the square of the dark matter density integrated over the line of sight for each of our assumed profiles.

\begin{figure*}
\mbox{\includegraphics[trim=203mm 200mm 200mm 160mm, clip, width=18.5cm, angle=180]{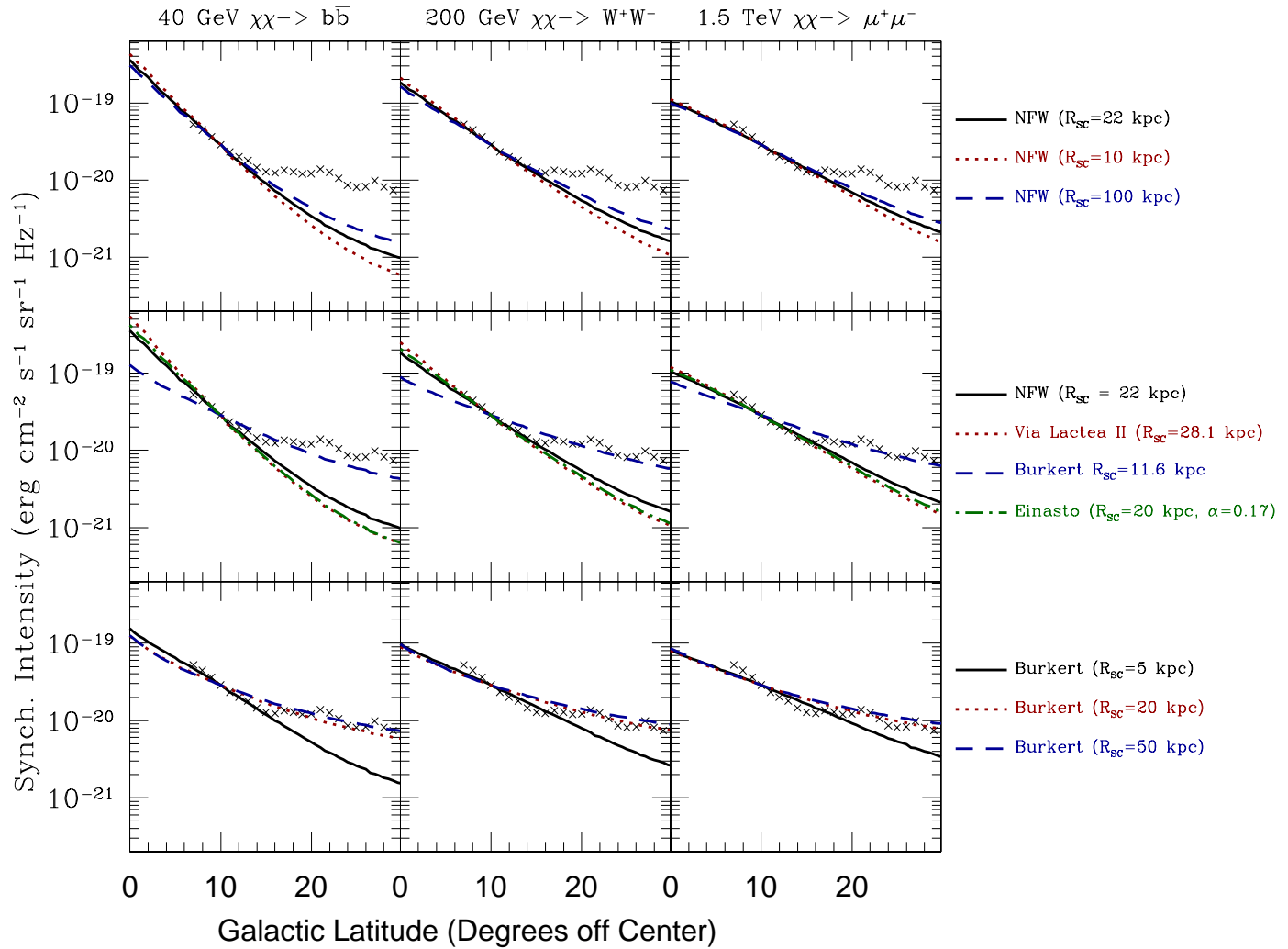}}
\caption{\it\small \label{profileplot} Simulated dark matter haze for three dark matter channels (40 GeV $\chi\chi$ -$>$ $b\bar{b}$ ({\bf Soft}, left), 200 GeV $\chi\chi$ -$>$ W$^{+}$W$^{-}$ ({\bf Wino}, center), 1500 GeV $\chi\chi$ -$>$ $\mu^{+}\mu^{-}$ ({\bf Hard}, right)) for illustrative choices of the dark matter density profile (top) and scale radius in the NFW profile (bottom). In each plot, the default simulation is shown in solid black.}
\end{figure*}

In Figure~\ref{profileplot}, we show the resulting dark matter haze for several choices of the NFW scale radius (top), various dark matter density profiles (middle), and several choices of the scale radius for the Burkert profile (bottom). As in previous plots, the results are normalized to the synchrotron haze at 10$^{\circ}$ latitude and at a frequency of 23 Ghz. Our default simulation is shown in black in the top two rows, but is not shown in the bottom row. 

We find that even large changes in the NFW profile scaling radius (top) have only a small effect on the morphology of the dark matter haze. For a scale radius of 100~kpc the ratio in synchrotron emission between 7$^\circ$ and 30$^\circ$ is still 37 for our {\bf Soft} spectrum, 21 for the {\bf Wino} spectrum, and 15 for the {\bf Hard} spectrum (as opposed to 6.7 for the WMAP observation). This again illustrates the point made in Figures~\ref{nodiffusion}~and~\ref{diffusionplot}~(top) - that the majority of synchrotron radiation seen at 30$^{\circ}$ is derived from leptons produced near the galactic center, which then diffuse to high latitudes.

This analysis is again confirmed upon examination of several dark matter density profiles (middle). The NFW, Via Lactea, and Einasto profiles all predict central regions with high dark matter density, and all three profiles result in very similar morphologies for the dark matter haze. This similarity comes despite the very differing DM densities near the galactic center in these profiles - with the average density throughout the inner 0.5~kpc of the Einasto profile being approximately twice as large as the Via Lactea profile and four times as large as the NFW profile. Since leptons produced in this region diffuse to high latitudes and dominate the dark matter synchrotron signal, these density profiles greatly affect the cross-section normalization in each model, but do not control it's morphology, which is governed instead by the diffusion setup. This allows us to infer that the addition of an additional even cuspier dark matter component into a profile with high central density will only significantly affect the assumed dark matter cross-section, with little effect, however, on the resultant synchrotron morphology.

However, the Burkert profile, which does not include a dense galactic center, gives a spatial variation which is much more consistent with the WMAP haze. Because the Burkert profile makes dark matter annihilation at high latitudes relatively more important compared to annihilations at the galactic center, it creates a much flatter distribution similar to the results shown in Figure~\ref{nodiffusion}. However, we note that removing the dense galactic center region eliminates the vast majority of our dark matter induced synchrotron radiation. Thus, to match the intensity of the WMAP haze with the Burkert profile, we require much higher normalization constants ($\sim$3010 for the {\bf Soft} dark matter model, $\sim$2350 for the {\bf Wino} model, and $\sim$37700 for the {\bf Hard} model)

As a result, to create a match to the WMAP haze through changes in the dark matter density profile, one might invoke a distribution like the Burkert profile which increases the importance of annihilations occurring at high galactic latitude\footnote{Alternatively, we note that a galactic substructure distribution with a broad radial distribution of subhalos might also effectively provide a boost at large radii which can mimic a similar DM density distribution to the cored Burkert profile}. In Row 3, we show the simulated dark matter haze for several values of the scale radius in the Burkert profile. Unlike the case of the NFW profile (top), the increased dependence on local dark matter annihilations in the Burkert profile creates a morphological dependence on the scale radius. For low scale radii like R$_{SC}$~=~5~kpc, we see a sharper cutoff that does not match the morphology of the WMAP haze. In contrast, we see very good matches for scaling radii between 11.6-50~kpc, depending on the employed dark matter particle setup. 

\section{The Haze Morphology: Role of Galactic Magnetic Fields}\label{sec:morphob}

A final important parameter in determining the spatial distribution of synchrotron radiation is the morphology of galactic magnetic fields. Changes in the galactic magnetic field morphology may produce reasonable matches for the WMAP haze because magnetic fields (1) allow for variations in the spatial morphology of synchrotron radiation without greatly affecting cosmic ray or $\gamma$-ray observations, and (2) are rather poorly constrained away from the solar position. However, we note that changes in the galactic magnetic field strength will not only affect the dark matter Haze, but will alter the intensity and morphology of all galactic synchrotron foregrounds, possibly eliminating or increasing the need for a WMAP haze. In this analysis, we artificially hold the observed WMAP haze constant through changes in the magnetic field intensity, as we only aim to show that known uncertainties in the galactic magnetic field could greatly alter the morphology of dark matter matter driven synchrotron emission.

\begin{figure*}
\mbox{\includegraphics[trim=203mm 203mm 200mm 70mm, clip, width=18.5cm, angle=180]{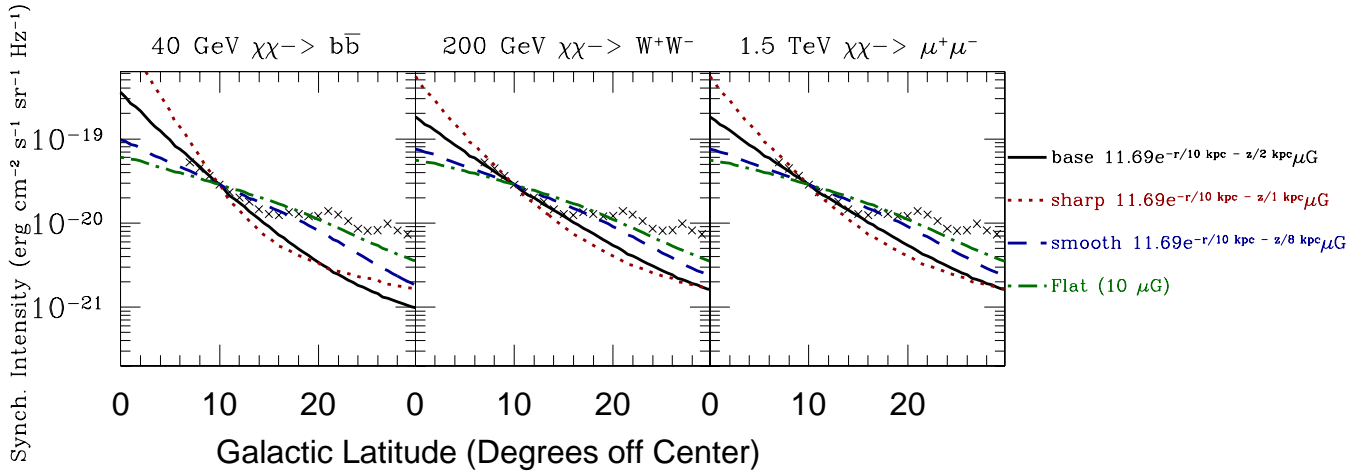}}
\caption{\it\small \label{bplot} Simulated dark matter haze for three dark matter channels (40 GeV $\chi\chi$ -$>$ $b\bar{b}$ ({\bf Soft}, left), 200 GeV $\chi\chi$ -$>$ W$^{+}$W$^{-}$ ({\bf Wino}, center), 1500 GeV $\chi\chi$ -$>$ $\mu^{+}\mu^{-}$ ({\bf Hard}, right)) for illustrative choices of the galactic magnetic field. In this plot, the default simulation parameters are shown in black with steeper and smoother magnetic fields shown in the various colors and line types.}
\end{figure*}

In Figure~\ref{bplot}, we evaluate the dark matter haze for several models of the magnetic field which follow the functional form of 
\begin{equation}
B=B_0\exp(-r/r_{0}~-z/z_0), 
\end{equation}
for various values for z$_0$. In addition, we model a flat (technically: scale radius of 99.9~kpc in both z and r) component equivalent to that employed in previous studies, e.g. \citet{Hooper:2007gi}. Again, the black line shows our base profile and we normalize all results to match the WMAP haze at 10$^\circ$ latitude. We find that employing a flat magnetic field profile greatly increases the synchrotron radiation at high latitudes above the galactic center producing intensity ratios of 10.5 ({\bf Soft}), 10.5 ({\bf Wino}), 9.35 ({\bf Hard}) between 7$^\circ$ and 30$^\circ$ in latitude. These values are more comparable to the ratio $\sim$~6.7 in the WMAP haze. However, we see that the flat magnetic field profile does not match the sharp rise in the WMAP haze between 7-10$^\circ$ in latitude and thus underproduces the haze in that region. The sharp magnetic field cutoff falls off very sharply at higher latitudes, while the smooth magnetic field model produces an intermediate case between the base and flat profiles. We note that several previous attempts at dark matter models of the WMAP haze (e.g. \cite{Hooper:2007kb}) use a magnetic field which is spatially uniform, and show results that are consistent with those shown here. 

Since the spatial variation of the magnetic field far from the solar neighborhood is highly uncertain, all magnetic fields presented here are fairly reasonable choices. This creates a large uncertainty in putative dark matter models fitting the WMAP haze, as the morphology of an additional synchrotron component can be adjusted by at least a factor of two from low to high latitudes. In addition to the freedom of choosing a normalization constant to match the intensity of the WMAP haze, this allows a secondary normalization constant which effectively controls the morphology of the haze as a function of latitude. This fact can be interpreted in two ways: either as evidence that dark matter models can successfully describe an excess synchrotron haze, or that the existence of a dark matter model which matches the morphology of the WMAP haze is not convincing proof of the models correctness, because uncertainties in the magnetic field make the underlying diffuse morphology and spectrum difficult to pinpoint.

\section{Towards a Self-Consistent Picture: Constraints from Cosmic Ray Data}
\label{sec:cosmicrays}

Until this point, we have simply made {\em ad hoc} changes to the diffusion, magnetic field, and dark matter density profiles without regard for constraints which may exist from other observations. In the remaining sections of this paper, we will much more closely examine the reasonability of our parameter space choices based on various other observable quantities. In particular, we outline the correlated ranges for the parameters influencing cosmic ray transport according to local constraints from cosmic ray data.

As seen in Figure~\ref{diffusionplot}, changes in both the diffusion constant and Alfv\'en velocity can greatly affect the morphology of the dark matter haze, potentially allowing well-motivated dark matter particle physics setups to match the observed WMAP haze. However, changes in the diffusion setup do not affect only the leptons produced by  dark matter annihilation, but affect the morphology of all cosmic ray species as well. Since cosmic rays interact as they propagate in the Galaxy, ratios of  secondary to primary cosmic ray species provide a valuable consistency check for the assumed cosmic ray diffusion models. We note that these same constraints do not exist for most changes in the dark matter profile or the magnetic field models, as neither have a large effect on the galactic primary to secondary ratios.

\begin{figure*}
\mbox{\includegraphics[width=1.0\textwidth,clip]{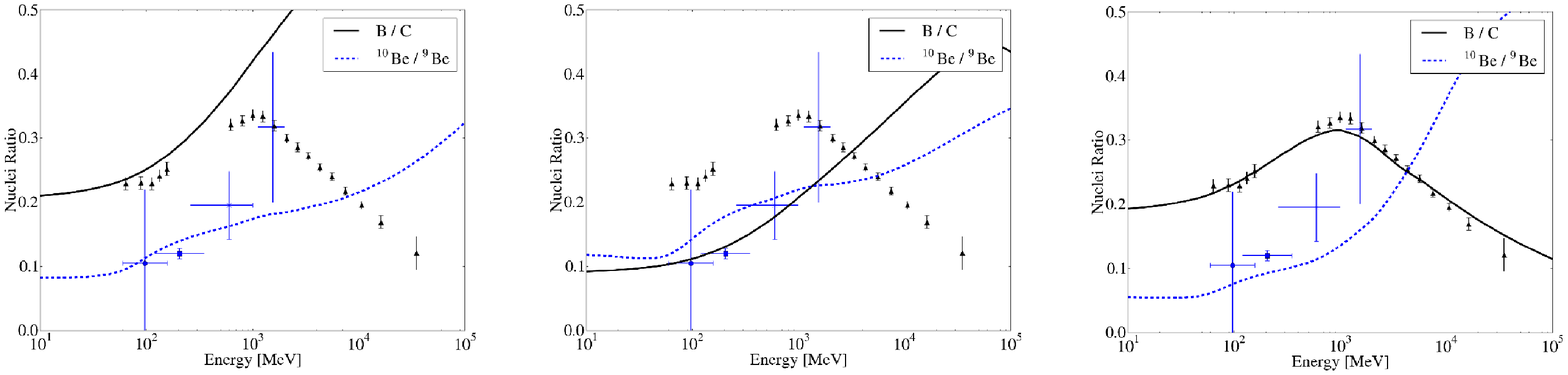}}
\caption{ \it\small \label{fig:nucleiplot} The Boron to carbon ratio (black solid / black triangles) and the $^{10}$Be to $^9$Be ratio (blue dashed / blue squares) modeled at the solar position for three important cosmic ray propagation setups: D$_0$~=~1~x~10$^{28}$~cm$^2$~s$^{-1}$ (left), v$_\alpha$ = 100 km s$^{-1}$ (middle), and our base model (right).}
\end{figure*}

In Figure~\ref{fig:nucleiplot}, we show some of the simulated cosmic ray data (the Boron to Carbon Ratio and the $^{10}$Be/$^{9}$Be ratio) for the two diffusion models in Section~\ref{sec:morpho}, which matched the morphology of the WMAP haze: (1) a low diffusion constant of D$_0$~=~1~x~10$^{28}$~cm$^2$~s$^{-1}$ (left) and (2) a high Alfv\'en velocity of v$_\alpha$ = 100 km s$^{-1}$ (middle)). We also show the nuclei fluxes for our default {\tt Galprop} model (right). We note that the large shifts in diffusion parameters necessary to match the WMAP haze also have large effects on the simulated primary to secondary fluxes and ratios. Both models are definitively ruled out by current observations at levels much greater than 10$\sigma$. In our default {\tt Galprop} model (which of course is refined in order to match observed cosmic ray data) we indeed find very close matches to current observations.

%%%%%%%%%%%%%%%%%%%%%%%%%%%%%%%%%%%%%%%%%%%%%%%%%%%
%%%%%%%%%%%%%%%%%%%%%%%%%%%%%%%%%%%%%%%%%%%%%%%%%%%

\begin{figure*}
\mbox{\includegraphics[width=1.0\textwidth,clip]{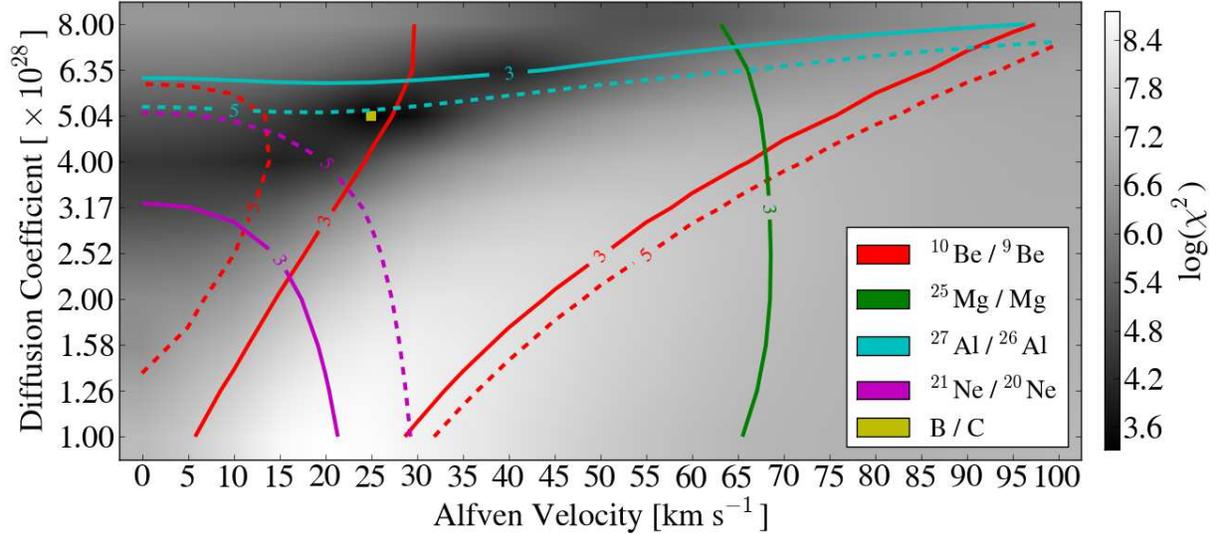}}
\caption{ \it\small \label{fig:nucleigrid} Constraints from Nuclei fluxes and primary to secondary ratios as a function of Alfv\'en velocity and diffusion constant. Darker areas indicate better fits to simulation data (lower $\chi ^2$). The minimum falls around a diffusion constant of 5.0~x~10$^{28}$~cm$^2$s$^{-1}$ and an Alfv\'en velocity of 25~km~s$^{-1}$. Colored lines for each ratio show 3 and 5 sigma constraints on the dataset \citep{Engelmann:1990zz,2001ApJ...563..768Y, 2005ApJ...634..351B, 2004ApJ...611..892H}.}
\end{figure*}

However, in our analysis thus far, we have evaluated only first order changes in each diffusion parameter, while the rest of our diffusion setup is kept constant. It is entirely possible that some combination of changes will create a match between the dark matter Haze and WMAP haze, while maintaining the observed primary/secondary ratios. In Figure~\ref{fig:nucleigrid}, we show the two dimensional parameter space spanned by a reasonable range for the diffusion constant and for the Alfv\'en velocity, which are two of the most significant parameters affecting the dark matter haze morphology. For each point in our parameter space, we calculate a total $\chi^{2}$ from available cosmic ray ratio data \citep{Engelmann:1990zz,2001ApJ...563..768Y, 2005ApJ...634..351B, 2004ApJ...611..892H}. The value of $\chi^{2}$ is calculated similarly to \citet[eq. 6]{Cumberbatch:2010ii}, but without reducing the total variance:

\begin{equation}
\chi^2 = \Sigma_j \Sigma_i^{Nj} \frac{(D_{ij} - T_{ij})^2}{\sigma^2_{ij}}
\end{equation}

where D$_{ij}$, $\sigma_{ij}$, and T$_{ij}$ are the central values, 1$\sigma$ errors, and predicted abundances respectively of the $i$th datapoint with the $j$th experimental dataset. This value is represented in the z-axis of Figure~\ref{fig:nucleigrid}.  Contour lines trace out the regions where each ratio is consistent with the data to within three and five sigma.  The tightest constraint comes from the Boron to Carbon ratio, which gives 3~$\sigma$ constraints limited to only a signal point in our gridding simulation, and 5~$\sigma$ constraints on the diffusion constant above 4.0~x~10$^{28}$~cm$^2$s$^{-1}$ and below 6.0~x~10$^{28}$~cm$^2$s$^{-1}$. The other ratios also appear to have reasonable values around that same point, with a diffusion constant of D$_0\sim$5.0~x~10$^{28}$~cm$^2$s$^{-1}$ and an Alfv\'en velocity of 25~km~s$^{-1}$. 

In Figure~\ref{dzhgrid}, we show a similar two-dimensional plot of the diffusion constant compared the height of the diffusion region. We indeed note a direct correlation between higher diffusion constants and diffusion heights, which creates a best fit region which is significantly more extended than the two dimensional fit of the diffusion constant with Alfven velocity. We find no constraints from either the Ne or Mg ratios, and do not include these in the fit. However, reasonable simulation heights rule out diffusion constants larger than approximately 1.5~x~10$^{29}$~cm$^2$s$^{-1}$. In Figure~\ref{dzhplot}, we see that usage of a higher diffusion constant along with a simulation height of 8~kpc does lead to a reasonable improvement in the morphology of the dark matter haze, compared to WMAP observations, especially for our {\bf Hard} annihilation spectra. However, even much larger simulation heights are unable to create a complete match to the haze, ruling out the possibility that this degeneracy condition alone is capable of explaining the morphological differences between the observed and simulated haze. 

We note that in this study we have only considered models employing diffusion parameters which are homogeneous throughout the diffusion zone. Physical intuition suggests that the high matter and radiation densities near the galactic center may locally decrease the diffusion constant and affect, for instance the re-acceleration of cosmic rays. While this may alter the cosmic ray primary-to-secondary ratios, our models indicate it should have little effect on the assumed dark matter synchrotron morphology - as in the cases of the NFW, Einasto, and Via Lactea profiles, the dark matter induced lepton population is entirely dominated by annihilations occuring near the galactic center. As long as changes in the diffusion constant are not strong enough to change this assumption, then the specific details of diffusion near the galactic center are degenerate with assumptions on the dark matter cross section (for changes in the diffusion constant), the input spectra of dark matter produced leptons (for changes in the reacceleration parameters) and the dark  matter density distribution in the galactic center region. In our models, the NFW, Einasto, and Via Lactea profiles all have very different central annihilation rates, and thus the morphological similarity between these models indicates that the synchrotron morphology is determined primarily by the diffusion setup outside of the galactic center, rather than by the details of the diffusion setup very near the galactic center.

In summary, our findings do not rule out the possibility that changes to the cosmic-ray propagation parameters compatible with local cosmic-ray data allow for significant effects on the high-energy lepton equilibrium distribution from a baseline setup. However, we showed that drastic changes to the morphology of the galactic $e^\pm$ equilibrium number density are produced by parameter choices that tend to create substantial issues in the resulting ratios of primary-to-secondary cosmic-ray fluxes. 

%%%%%%%%%%%%%%%%%%%%%%%%%%%%%%%%%%%%%%%%%%%%%%%%%%%
%%%%%%%%%%%%%%%%%%%%%%%%%%%%%%%%%%%%%%%%%%%%%%%%%%%

%%%%%%%%%%%%%%%%%%%%%%%%%%%%%%%%%%%%%%%%%%%%%%%%%%%
\begin{figure*}
\mbox{\includegraphics[width=1.0\textwidth,clip]{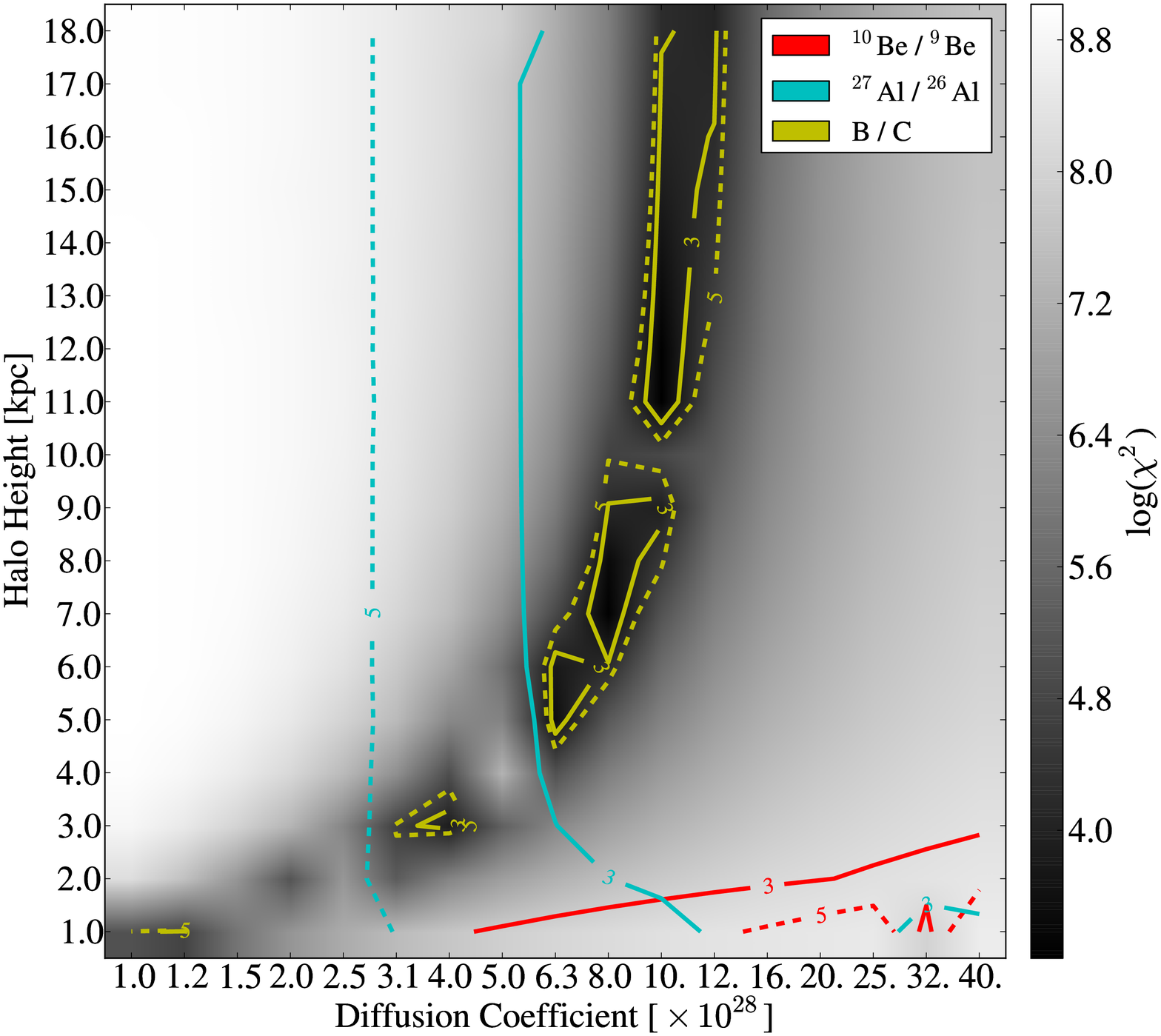}}
\caption{ \it\small \label{fig:dzhgrid} Constraints from Nuclei fluxes and primary to secondary ratios as a function of the simulation height and diffusion constant. Darker areas indicate better fits to simulation data (lower $\chi ^2$). A degeneracy exists along a line of constant diffusion constant divided by height. Colored lines for each ratio show 3 and 5 sigma constraints on the dataset. Some local maxima which are non-physical have been smoothed over.}
\end{figure*}
%%%%%%%%%%%%%%%%%%%%%%%%%%%%%%%%%%%%%%%%%%%%%%%%%%%

\section{Frequency dependence of dark matter Haze}\label{sec:radio}

\begin{figure*}
\mbox{\includegraphics[trim=203mm 200mm 200mm 70mm, clip, width=18.5cm, angle=180]{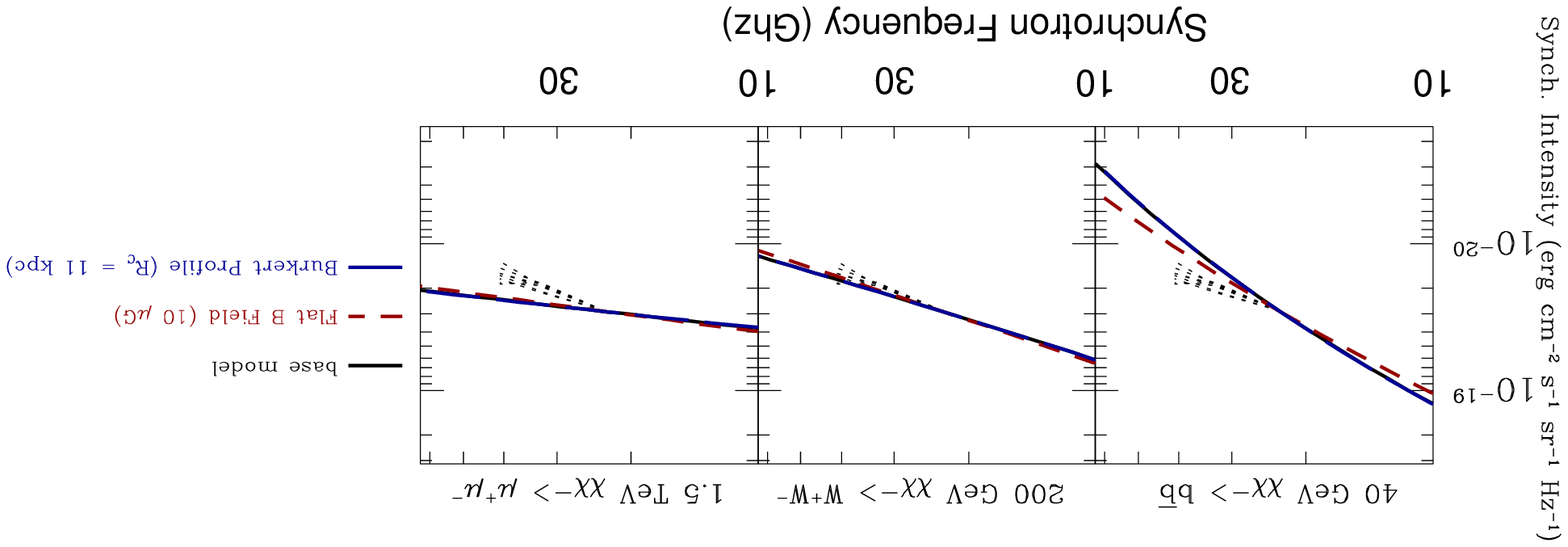}}
\caption{\it\small \label{fig:freqplot} Frequency dependence of the simulated dark matter haze for three dark matter decay channels (40 GeV $\chi\chi$ -$>$ $b\bar{b}$ ({\bf Soft}, left), 200 GeV $\chi\chi$ -$>$ W$^{+}$W$^{-}$ ({\bf Wino}, center), 1500 GeV $\chi\chi$ -$>$ $\mu^{+}\mu^{-}$ ({\bf Hard}, right)) at 10 degrees below the galactic plane for our default and two models which provide reasonable WMAP haze matches: (1) base model (solid black), (2) Flat Magnetic Field (red dashed), (3) Burkert profile (blue long-dashed). The dotted region shows a range of WMAP frequency dependences for each degree in between 7-15 degrees below the galactic plane. All values have been normalized to match at 23 Ghz.}
\end{figure*}

One further test of the dark matter interpretation of the WMAP haze, originally proposed by \citet{2006PhRvD..73b3521A}, concerns examining the frequency dependence of simulated dark matter haze models. The main parameter space uncertainties for these models result from the dark matter annihilation pathway and the magnetic field model, which can produce either harder or softer spectra. In Figure~\ref{fig:freqplot}, we show the spectra of our default model, as well as a model employing a Burkert profile and the flat magnetic field model from Section~\ref{sec:cosmicrays}. The simulations are shown at 10$^\circ$ below the galactic center and normalized to match the observed data at 23~Ghz. The data shown here originates from all latitudes between 7-15 degrees, and averaged over the longitudinal strip -10$^\circ$~$<$~$l$~$<$~10$^\circ$ and are normalized at 23~Ghz to match the value at 10$^\circ$ below the galactic center. We show WMAP data at only 23 Ghz, 33 Ghz, and 41 Ghz, where the observed error bars do not dominate the residual signal.

We point out two important results:
\begin{enumerate}
\item the dominant contribution to the frequency dependence of the dark matter haze relates to the mass and annihilation pathway of the dark matter particle; Our three illustrative dark matter models show very different frequency dependences, which can help differentiate between each model, and 
\item the {\bf Wino} model provides the best match for the spectral characteristics of the WMAP haze. 
\end{enumerate}
We note, however, that we have not rigorously examined either the systematic or observational errors in this measurement, and thus further study is needed to determine whether these frequency tests rigorously rule out any of the models presented here. Future measurements from the Planck satellite can be expected to provide an improved spectral analysis, making this technique and important multi-wavelength test of any future residual signal.

\section{Comparison to Indirect Detection on other energy scales}\label{sec:invco}

\begin{figure*}
\mbox{\includegraphics[trim=203mm 200mm 200mm 160mm, clip, width=18.5cm, angle=180]{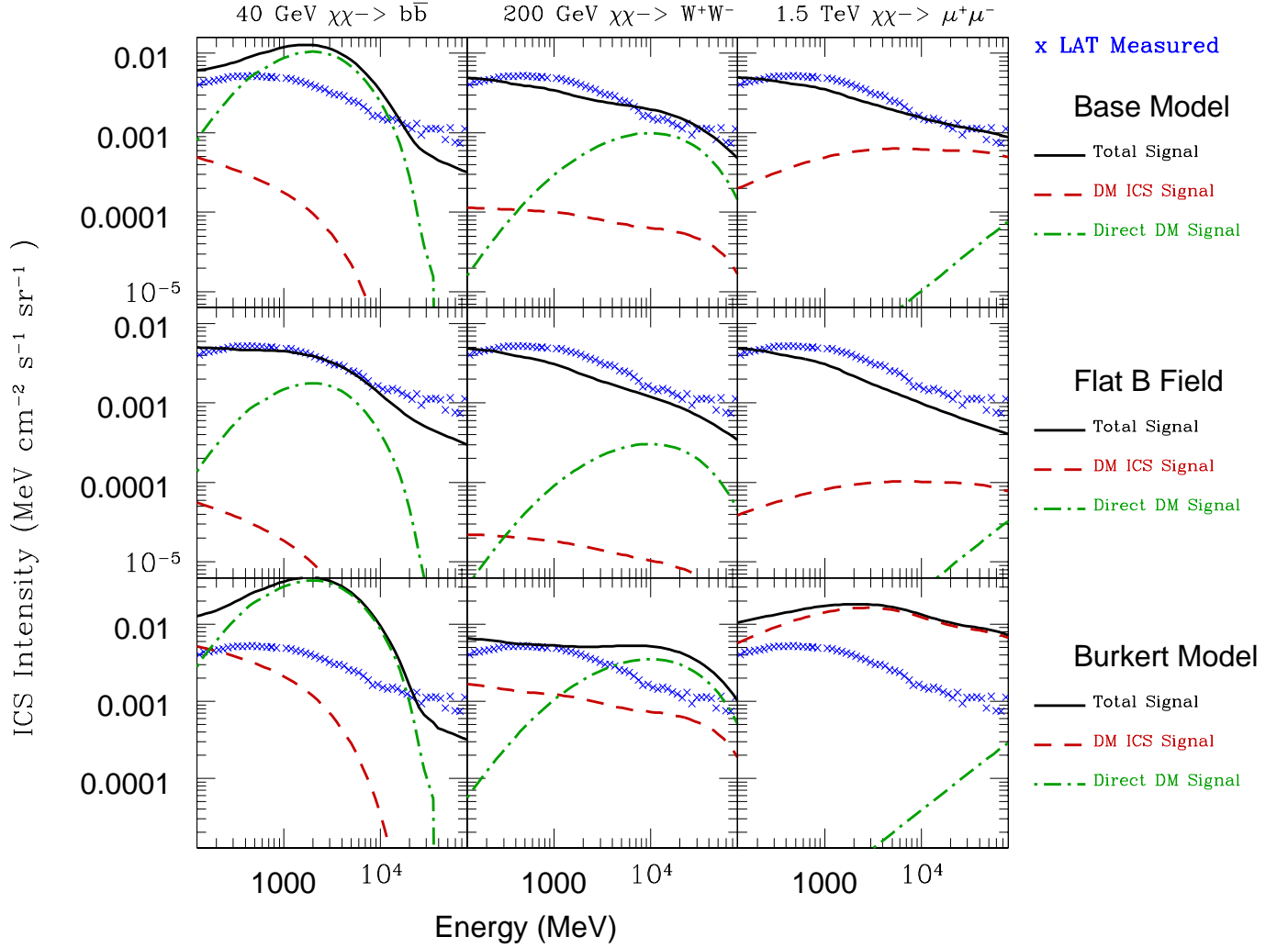}}
\caption{\it\small \label{fig:icshaze} Expected Fermi Signal from each decay pathway for the base model (top), a flat B-field of 10~$\mu$G (middle), and a model employing a Burkert profile with R$_{sc}$~=~11.6~kpc bottom, and shown for three dark matter decay channels (40 GeV $\chi\chi$ -$>$ $b\bar{b}$ ({\bf Soft}, left), 200 GeV $\chi\chi$ -$>$ W$^{+}$W$^{-}$ ({\bf Wino}, center), 1500 GeV $\chi\chi$ -$>$ $\mu^{+}\mu^{-}$ ({\bf Hard}, right)). The red dash shows the expected Inverse Compton (ICS) emission based on the {\tt Galprop} model for the galactic ISRF, while the green dot-dashed line is based on the direct $\gamma$-ray production from a {\tt DarkSUSY} model, and normalized based on the input dark matter density profile. The total emission (black solid) includes {\tt Galprop} based emission for $\pi^0$-decay and bremsstrahlung emission, as well as power law models for the extragalactic background (Energy$^{-2.45}$ normalized to 1.1 x 10$^{-3}$ MeV~cm$^{-2}$~s$^{-1}$~sr$^{-1}$ at 1~GeV) and point sources  (Energy$^{-2.4}$ normalized to 3.1 x 10$^{-4} $ MeV~cm$^{-2}$~s$^{-1}$~sr$^{-1}$ at 1~GeV). The observed Fermi-LAT $\gamma$-ray flux is obtained from \citep{Abdo:2009mr}.}.
\end{figure*}

While we have focused exclusively on the detectability of the synchrotron contribution of dark matter, we note that \emph{any} additional primary lepton flux (be it from an annihilating dark matter candidate, pulsars, SN remnants etc.), is guaranteed to produce not only \emph{some} synchrotron signal, but also signals at GeV energies due to inverse Compton scattering off of the interstellar radiation field of the Galaxy. We can thus put further constraints on our dark matter models of the WMAP haze by evaluating the expected $\gamma$-ray signature of each setup and enforcing constrains from recent Fermi data. We focus here on three morphologically interesting models: (1) our base setup (top), (2) a model with a flat magnetic field (middle), and (3) a model employing a Burkert dark matter density profile with R$_{C}$~=11.6~kpc (bottom). In Figure~\ref{fig:icshaze}, we plot the expected Fermi signal due in each decay pathway due to both direct $\gamma$-ray production (green dot-dash), and inverse-Compton emission from scattering off of the galactic ISRF (red dash). For the total signal (black solid), we add emission from $\pi^0$ decay and bremsstrahlung emission from our default {\tt Galprop} models, as well as power law spectra for the extragalactic isotropic background (Energy$^{-2.45}$) and point sources (Energy$^{-2.4}$) (from inspection of \citep{Abdo:2009mr}). 

We find that several of these models significantly over-predict the observed Fermi $\gamma$-ray signal at intermediate latitudes. For our {\bf Soft} and {\bf Wino} models, we find that direct $\gamma$-ray production is the dominant DM signal at intermediate latitudes. The large boost factors necessary to recreate the intensity of the WMAP haze overproduce the observed Fermi signal in both our base and Burkert models, while the stronger magnetic field in our Flat B model allows the $\gamma$-ray emission to match observations. For our {\bf Wino} models, we note an overproduction of direct $\gamma$-ray signals from our Burkert model, while the base and flat magnetic field model both yield reasonable results. 

In the case of our {\bf Hard} models, we find the dominant $\gamma$-ray production mechanism to stem from ICS of the ISRF. This signal overproduces the observed Fermi signal in our Burkert model, while it provides reasonable fits in our base and flat magnetic field models. 

Stronger constraints also come from searches for dark matter annihilation in dwarf galaxies and clusters. The results of \citet{2010ApJ...712..147A} (Figure 3, top) allows us to place a cross-section limit of approximately 8~x~10$^{-26}$~cm$^3$~s$^{-1}$ on $b\bar{b}$ annihilation spectra. This definitively rules out our {\bf Soft} default model as it requires an annihilation cross-section of approximately 1.1~x~10$^{-24}$~cm$^3$~s$^{-1}$. However, in the wino case (Figure 3, bottom), the best Fermi limits from dwarfs give a cross-section of approximately 1~x~10$^{-23}$~cm$^3$~s$^{-1}$, which is well above our model {\bf Wino} cross section of 8.3~x~10$^{-25}$~cm$^3$~s$^{-1}$. In the case of our {\bf Hard} spectrum, the best limits lie at approximately 5~x~10$^{-22}$~cm$^3$~s$^{-1}$, which is well above our necessary cross-section of 1.2~x~10$^{-23}$~cm$^3$~s$^{-1}$.

We also note several other relevant constraints on the ICS emission. Work by \citet{Papucci:2009gd} (Figure 3, NFW Profile) places limits of 4~x~10$^{-25}$~cm$^3$~s$^{-1}$ on our {\bf Soft} model, 1.3~x~10$^{-24}$~cm$^3$~s$^{-1}$ on our {\bf Wino} model, and 5~x~10$^{-22}$~cm$^3$~s$^{-1}$ on our {\bf Hard} model - which gives better constraints on our wino model. Fermi models of the isotropic background \citep{Abdo:2010dk} (Figure 5, MSII-Sub1 limit) give 95\% limits of 5~x~10$^{-25}$~cm$^3$~s$^{-1}$ on our {\bf Soft} model. 

As intuition suggests, the necessary cross-sections always decrease when we employ a stronger magnetic field model. In the case of our {\bf Soft} annihilation spectrum, the flat magnetic field model lowers the necessary cross-section to just 1.83~x~10$^{-25}$~cm$^3$~s$^{-1}$, which lies closer to the constraints imposed by Fermi observations. This yields the enticing possibility that low mass dark matter annihilation spectra with intensities able to match the haze will either be confirmed or ruled out by Fermi in the very near future. 

Notice that further constraints on dark matter annihilation and its emission from the galactic center region can be obtained by studying regions very close (on the order of a few arcsec) to Sag A*, under educated assumptions on the magnetic field intensity in that region \cite{Regis:2008ij}; in particular, synchrotron emission is severely constrained by both radio and X-ray observations at small angular scales around Sag A* \cite{Regis:2008ij}. Given the coarse spatial grid we employ in this study, and the fact that we are concerned with much larger angular scales, we do not discuss this type of constraints here.

\section{Discussion and Conclusions}\label{sec:disc}

WIMP annihilation in the Galaxy unavoidably yields high-energy electrons and positrons (barring the possibility of pair annihilations in neutrino pairs or particles outside the Standard Model). In turn, this implies a diffuse galactic emission, whose morphology is affected by both the distribution of dark matter and charged cosmic-ray transport in the Galaxy. In this study, we outlined in detail how the form and nature of the WIMP radio halo depends upon the particle model for dark matter as well upon astrophysical parameters such as the galactic magnetic fields and the parameters entering the propagation of cosmic rays. We pointed out how cosmic-ray data very significantly constrain the morphology of the WIMP radio haze, and pointed out the importance of other secondary emission channels, in particular Inverse Compton.

Given the interest in a possible excess emission at radio frequencies -- the WMAP haze -- we have shown that standard dark matter annihilation pathways and cosmic ray diffusion models do not provide a reasonable match for the morphology of the excess. Dark matter density profiles with high densities near the galactic center produce synchrotron morphologies which are dominated by the diffusion of charged leptons away from the galactic center. However, changes in the diffusion setup do not allow one to reproduce the morphology of the WMAP haze while remaining consistent with local observations of cosmic ray primary to secondary ratios. Models which lack a dense dark matter distribution in the galactic center (e.g. the Burkert profile) require large cross-section enhancements which are ruled out by Fermi observations. 

We also found that the particle physics properties of dark matter models which seek to explain both the WMAP haze as well as higher energy observations such as those from the Fermi-LAT are beginning to be constrained. Low energy dark matter decay spectra such as our {\bf Soft} model are unable to reproduce the intensity of the WMAP haze without overproducing the Fermi signal with $\gamma$-rays produced via direct annihilation.

However many uncertainties still exist in the magnetic field morphology to allow viable matches for the observed excess in synchrotron emission. The magnetic field morphology primarily influences the synchrotron intensity and morphology, with very little effect on $\gamma$-ray or cosmic ray constraints. The best constraints on galactic magnetic fields would come from careful examination of the WMAP dataset which would determine the morphology of any WMAP haze resulting from various magnetic field models. Such an analysis is beyond the scope of this study, but has been examined in both \citet{Gold:2010fm} and \citet{Dobler:2007wv}, with contrasting results. Furthermore, many parameters not mentioned in this work may be fine tuned to produce a reasonable match to all known constraints. However, the formidable size of this parameter space makes the existence of such matches unconvincing evidence that the WMAP haze arises from a dark matter source.

Further inquiry in the nature of an excess diffuse radio emission is certainly warranted, especially since e.g. any additional IC signal  in the high energy sky \citep{Dobler:2009xz,Linden:2010ea} implies the production of some microwave residual detectable in the microwave sky. Diffuse radio emission, and constraints on it, can thus be regarded as a powerful multi-wavelength diagnostic of any claims of new anomalous signals. Lastly, future data collected by the Planck telescope \cite{Bouchet:2009tr} will be imperative in determining the intensity of the WMAP haze, and in extending its spectrum to the higher energy regime.

\begin{acknowledgments}
\noindent We thank Troy Porter for many helpful conversations involving {\tt Galprop} models, and Gregory Dobler and Douglas Finkbeiner for providing observational models of the WMAP Haze. TL is supported by a GAANN Fellowship from the Department of Education. SP is partly supported by an Outstanding Junior Investigator Award from the US Department of Energy and by Contract DE-FG02-04ER41268, and by NSF Grant PHY-0757911. BA acknowledges support from NASA GI Grant NNX08AV59G.
\end{acknowledgments}

\bibliography{haze}

\end{document}